\documentclass[times]{elsarticle}

 \pdfoutput=1

\usepackage{bm}
\usepackage{amssymb}
\usepackage{amsmath}

\usepackage{lineno}
\modulolinenumbers[5]

\usepackage{color}

\begin{document}

\def\tc{$T_{\mathrm{c}}$\ }
\def\t6as{$\mathrm{(TMTSF)_{2}AsF_{6}}$\ }
\def\tmx{(TMTSF)$_{2}$(ClO$_{4}$)$_{(1-x)}$(ReO$_{4}$)$_x$\ }
\def\tmxns{(TMTSF)$_{2}$(ClO$_{4}$)$_{(1-x)}$(ReO$_{4}$)$_x$}
\def\tmc{$\mathrm{(TMTSF)_{2}ClO_{4}}$\ }
\def\tmcns{$\mathrm{(TMTSF)_{2}ClO_{4}}$}
\def\tms{$\mathrm{(TMTSF)_{2}AsF_{6(1-x)}SbF_{6x}}$\ }
\def\tmps{$\mathrm{(TMTTF)_{2}PF_{6}}$\ }
\def\tmttfsbf6{$\mathrm{(TMTTF)_{2}SbF_{6}}$\ }
\def\tmttfsbfns{$\mathrm{(TMTTF)_{2}SbF_{6}}$}
\def\tmttfasf6{$\mathrm{(TMTTF)_{2}AsF_{6}}$\ }
\def\tmttfbf4{$\mathrm{(TMTTF)_{2}BF_{4}}$\ }
\def\tmttfbr{$\mathrm{(TMTTF)_{2}Br}$\ }
\def\tmttfbrns{$\mathrm{(TMTTF)_{2}Br}$}
\def\tmtsfreo4{$\mathrm{(TMTSF)_{2}ReO_{4}}$\ }
\def\tmno3{$\mathrm{(TMTSF)_{2}NO_{3}}$\ }
\def\tm2x{$\mathrm{(TM)_{2}X}$\ }
\def\tm2xns{$\mathrm{(TM)_{2}X}$}
\def\tq{TTF-TCNQ\ }
\def\tqns{TTF-TCNQ}
\def\tsq{$\mathrm{TSeF-TCNQ}$\ }
\def\qnq{$(Qn)TCNQ_{2}$\ }
\def\R{$\mathrm{ReO_{4}^{-}}$}  
\def\C{$\mathrm{ClO_{4}^{-}}$}
\def\P{$\mathrm{PF_{6}^{-}}$}
\def\tqr{$\mathrm{TCNQ^\frac{\cdot}{}}$\ }
\def\nmpq{$\mathrm{NMP^{+}(TCNQ)^\frac{\cdot}{}}$\ }
\def\q{$\mathrm{TCNQ}$\ }
\def\nmp{$\mathrm{NMP^{+}}$\ }
\def\f{$\mathrm{TTF}\ $}
\def\pc{$P_{\mathrm{c}}$\ }
\def\pcns{$P_{\mathrm{c}}$}
\def\hc2{$H_{\mathrm{c2}}$\ }
\def\nmq{$\mathrm{(NMP-TCNQ)}$\ }
\def\ts{$\mathrm{TSF}$}
\def\tsm{$\mathrm{TMTSF}$\ }
\def\tst{$\mathrm{TMTTF}$\ }
\def\tmp6{$\mathrm{(TMTSF)_{2}PF_{6}}$\ }
\def\tmpns{$\mathrm{(TMTSF)_{2}PF_{6}}$}
\def\tms2x{$\mathrm{(TMTSF)_{2}}X$}
\def\tm2x{$\mathrm{(TM)_{2}}X$\ }
\def\as{$\mathrm{AsF_{6}}$}
\def\sb{$\mathrm{SbF_{6}}$}
\def\pf{$\mathrm{PF_{6}}$}
\def\re{$\mathrm{ReO_{4}}$}
\def\ta{$\mathrm{TaF_{6}}$}
\def\cl{$\mathrm{ClO_{4}}$}
\def\no{$\mathrm{NO_{3}}$}
\def\4fb{$\mathrm{BF_{4}}$}
\def\ttdm{$\mathrm{(TTDM-TTF)_{2}Au(mnt)_{2}}$}
\def\edt{$\mathrm{(EDT-TTF-CONMe_{2})_{2}AsF_{6}}$}
\def\tfx{$\mathrm{(TMTTF)_{2}X}$\ }
\def\tsx{$\mathrm{(TMTSF)_{2}}X$\ }
\def\ttftcnq{TTF-TCNQ\ }
\def\ttf{$\mathrm{TTF}$\ }
\def\tcnq{$\mathrm{TCNQ}$\ }
\def\bedtttf{$\mathrm{BEDT-TTF}$\ }
\def\reo4{$\mathrm{ReO_{4}}$}
\def\bedtttfreo4{$\mathrm{(BEDT-TTF)_{2}ReO_{4}}$\ }
\def\et2i3{$\mathrm{(ET)_{2}I_{3}}$\,}
\def\et2x{$\mathrm{(ET)_{2}X}$\,}
\def\ket2x{$\mathrm{\kappa-(ET)_{2}X}$\,}
\def\cuncnbr{$\mathrm{Cu(N(CN)_{2})Br}$\,}
\def\ket2x{$\mathrm{\kappa-(ET)_{2}X}$\,}
\def\cuncncl{$\mathrm{Cu(N(CN)_{2})Cl}$\,}
\def\cuncs{$\mathrm{Cu(NCS)_{2}}$\,}
\def\betsfecl4{$\mathrm{(BETS)_{2}FeCl_{4}}$\,}
\def\bets{$\mathrm{BETS}$\,}
\def\et{$\mathrm{ET}$\,}

\newcommand{\sub}[1]{\ensuremath{_{\mathrm {#1}}}}
\newcommand{\sps}[1]{\ensuremath{^{\mathrm {#1}}}}

\newcommand{\vF}{\bm{v}\sub{F}}
\newcommand{\kF}{\bm{k}\sub{F}}
\newcommand{\kFu}{\bm{k}\sub{F\uparrow}}
\newcommand{\kFd}{\bm{k}\sub{F\downarrow}}
\newcommand{\vs}{\bm{v}\sub{s}}
\newcommand{\etal}{\textit{et~al.}}
\renewcommand{\deg}{^{\circ}}
\newcommand{\qFFLO}{\bm{q}\sub{FFLO}}
\newcommand{\Tc}{T\sub{c}}
\newcommand{\Hcc}{H\sub{c2}}
\newcommand{\dash}{^{\prime}}
\newcommand{\Tco}{T\sub{c}\sps{onset}}
\newcommand{\cstar}{c^{\ast}}
\newcommand{\bstar}{b^{\ast}}
\newcommand{\Tcz}{T\sub{c0}}



\begin{frontmatter}


\title{Novel superconducting phenomena in quasi-one-dimensional Bechgaard salts}

\author[label1]{Denis J\'erome\corref{cor1}}
\address[label1]{Laboratoire de Physique des Solides, CNRS UMR 8502, Universit{\'{e}} Paris-Sud, F-91405 Orsay, France}

\author[label2]{Shingo Yonezawa}
\address[label2]{Department of Physics, Graduate School of Science, Kyoto University, Kyoto 606-8502, Japan}

\cortext[cor1]{corresponding author}

\begin{abstract}It is the saturation of the transition temperature \tc in the range of 24~K for known materials in the late sixties which triggered the search for additional materials offering new coupling mechanisms leading in turn to higher $\Tc$'s.  As a result of this stimulation, superconductivity in organic matter was discovered in tetramethyl-tetraselenafulvalene-hexafluorophosphate, \tmpns, in 1979,  in the laboratory founded at Orsay by Professor Friedel and his colleagues in 1962. Although this  conductor is a prototype example  for low-dimensional physics, we mostly focus in this article on the superconducting phase of the ambient-pressure superconductor \tmcns, in which the superconducting phase has been studied most intensively among the TMTSF salts. 
We shall present a series of experimental results supporting nodal $d$-wave symmetry for the superconducting gap in these prototypical quasi-one-dimensional conductors.
\end{abstract}

\begin{keyword}
\texttt{One dimensional conductors, Organic superconductivity, Bechgaard salts, \tmc}
\end{keyword}

\end{frontmatter}


\section{Introduction --- Historical overview}

Searching for new materials exhibiting the highest possible values for the superconducting (SC) critical temperature \tc was  a  strong motivation in materials science in the early 70's, and the term ``high temperature superconductor'' was already commonly used referring
to the intermetallic compounds of the A15 structure, namely materials such as $\mathrm{Nb_{3}Sn}$ or $\mathrm{V_{3}Si}$~\cite{Hardy54}.

Extending the very successful explanation of the isotope effect in the Bardeen-Cooper-Schrieffer (BCS) theory, other models were proposed in which excitations of the lattice responsible for the electron pairing had been replaced by higher-energy excitations, namely, electronic excitations, with the hope of  finding new  materials with  \tc higher than those explained by the BCS theory.
The small electronic mass
$m_e$ of the polarizable medium would lead  to an enhancement of
\tc of the order of  ($M/m_e)^{1/2}$ times the value which is observed  in a conventional superconductor where $M$ is an atomic mass. This is admittedly a huge factor. V.~L.~Ginzburg~\cite{Ginzburg64a,Ginzburg64b} considered in 1964 the possibility for the pairing of electrons in
metal layers sandwiched between polarizable dielectrics through virtual excitations at high energy. 
But, the most provocative suggestion came from W.~A.~Little in 1964~\cite{Little64,Little65}, who predicted room-temperature superconductivity with a new pairing mechanism leading to a drastic enhancement of the superconducting $\Tc$.

The idea of Little
was rooted in the extension of the isotope effect proposed by BCS, replacing the  mediating phonon by an electronic excitation in especially designed quasi-one-dimensional (Q1D) macromolecules. However a prerequisite to the model of Little was the achievement of metallic conduction in organic molecular crystals. This was  not a trivial problem in the sixties.

A short time later, the synthesis of the first stable organic compound displaying metallic conduction below room temperature, the charge transfer complex \ttftcnq came out. This compound is made up of two
kinds of flat molecules each forming  segregated parallel conducting stacks. It fulfills the conditions for an organic conductor
as the orbitals involved in the conduction ($\pi$-HOMO, highest occupied molecular orbital and $\pi$-LUMO, lowest unoccupied molecular orbitals for \ttf and \tcnq respectively) are associated with the molecule as a whole rather than with a particular atom. Free carriers within each stack are given by an interstack charge  transfer \textit{at variance} with other organic conductors known at that time such as the conducting polymers, in which charges are provided by doping~\cite{Shirakawa77}. However, the conducting behaviour in \ttftcnq is stopped at low temperature by  a metal-insulator transition accompanying a Peierls distortion~\cite{Denoyer75}.  The Peierls ground state turned out to be very robust despite numerous attempts  to suppress it under  high pressure  making the one-dimensional (1D) conductor more two dimensional (2D)~\cite{Horovitz75,Jerome82}.
After more than thirty years, the insulating state is found to be almost suppressed at pressure as high as 8~GPa~\cite{Yasuzuka2007.JPhysSocJpn.76.033701}.


The clue to overcome the natural tendency for a 1D conductor to undergo a Peierls transition towards an  insulating ground state came after a fair amount of experimental works in physics together with chemistry using the newly discovered organic donor tetramethyl-tetraselena-fulvalene \tsm~\cite{Bechgaard79}.

The Copenhagen group led by Klaus Bechgaard, very experienced with the chemistry of selenium, succeeded in the
synthesis of a new series of conducting salts all based on the  TMTSF  molecule with the stoichiometry 2:1  namely, \tms2x,   where $X$ is an inorganic
mono-anion with various possible symmetry, octahedral (\pf, \as, \sb, \ta), tetrahedral (\4fb, \cl, \re )
 or triangular $\mathrm{(NO_3}$)~\cite{Bechgaard79}. All these compounds but the one with $X={}$\cl\ did exhibit an insulating ground state under ambient pressure. 

What is  so special with \tmpns, the prototype of the so-called  Bechgaard salts, unlike previously investigated \tqns, is the magnetic origin of the ambient-pressure insulating state~\cite{Andrieux81} contrasting with the Peierls-like ground states discovered previously in charge transfer compounds. The ground state of \tmpns\ turned out to be a spin density wave (SDW) state as shown in Fig.~\ref{molecule,structure,supra2}, similar to the predictions made by Lomer~\cite{Lomer62} in 1962 and by Overhauser~\cite{Overhauser60a} for metals.  However, the SDW has been suppressed under a hydrostatic pressure of about 9 kbar enabling the stabilization of metal-like conduction down to liquid helium temperature, and finally the stabilization of superconductivity below 1K found back in December 1979~\cite{Jerome80}, as presented in Fig.~\ref{PF6supra2.pdf}.

Soon after the discovery of superconductivity, it was revealed that the electronic band of \tms2x can be well modeled with the tight-binding model between TMTSF molecular sites~\cite{Grant1983PhysRevLett,Grant1983.JPhys.44.C3-847,Yamaji1984,Ducasse1985.JPhysC.18.L947,Ducasse1987,Pevelen2001EurPhysJB}. These tight-binding bands agree surprisingly well with recent first-principles band calculations~\cite{Ishibashi1999JPhysCondensMatter,Nagai2011.PhysRevB.83.104523,Nakamura2013.PhysRevB.88.125128,Alemany2014.PhysRevB.89.155124,Aizawa2015.unpublished}.
Moreover, it is now established that the Q1D electron gas model with weak-coupling limit explains fairly well key properties of the SDW phases in \tsx materials: both the suppression of the SDW phase under pressure~\cite{Yamaji82,Montambaux88} and the stabilization of magnetic-field-induced SDW phases~\cite{Gorkov84,Heritier84}. The non-interacting part of the Q1D electron gas model is defined in terms of a strongly anisotropic  electron spectrum yielding  an orthorhombic variant of the real open Fermi surface in the $ab$ plane of  the Bechgaard salts.  The spectrum $E(\bm{k}) = v_F(|k|-k\sub{F}) -2t_{b}\cos k_{b} - 2t_{b}'\cos 2k_{b}$  as a function of the momentum $\bm{k}=(k,k_{b})$  is characterized by an intrachain or longitudinal Fermi energy $E\sub{F}=v\sub{F}k\sub{F}$,  which takes a value around 3000~K in (TMTSF)$_2X$~\cite{Ducasse86,Pevelen2001EurPhysJB}; here $v_F$ and $k_F$ are the longitudinal Fermi velocity and wave vector. This energy is much larger than the interchain hopping
integral $t_{b}$ ($\approx 200$~K), in turn much bigger than the second-nearest neighbor transverse hopping amplitude $t_{b}'$. The latter stands as the antinesting parameter of the spectrum, which simulates the main influence of pressure in the model.

The unnesting parameters of the band structure $t^{'}_{b}$ and similarly $t^{'}_{c}$  for the $\cstar$ direction both play an important role  in the $T-P$ and $T-P-H$ phase diagrams of \tms2x. 
When $t^{'}_{b}$ exceeds a critical unnesting band integral of the order of the SDW transition temperature for the complete nesting ($\approx 15-30$~K)~\cite{Yamaji82,Montambaux88}, the SDW ground state is suppressed in favour of a metallic phase with the possibility of restoration of SDW phases under magnetic field along the $\cstar$ axis~\cite{Ishiguro98}.

The close proximity between antiferromagnetism and SC ground states of \tm2x (TM = TMTSF or TMTTF) superconductors and the deviation of the metallic phase from the traditional Fermi-liquid behaviour have been recognized as early as in the beginning of the eighties. The possibility for a pairing mechanism involving carriers on neighbouring chains in these Q1D conductors avoiding  the Coulomb repulsion  has been  proposed  by V. Emery in the context of the exchange phonon mechanism~\cite{Emery83}. Soon after, Emery and coworkers introduced the possibility that antiferromagnetic fluctuations play a role in the pairing mechanism~\cite{Emery86,Beal86} but concluded that superconductivity could not emerge from pairing on the same organic chain. The exchange of spin fluctuations between carriers on neighbouring chains was thus proposed~\cite{Emery86} to provide the necessary glue for pairing in analogy with  the exchange of charge density waves proposed by Kohn and Luttinger~\cite{Kohn65} in the context of a new pairing mechanism in low dimensional conductors.

In the context of  superconductivity in heavy fermions metals discovered the same year as  organic superconductivity~\cite{Steglich79}, J. Hirsch  performed a Monte Carlo simulation of the Hubbard model.
He showed an enhancement of anisotropic spin-singlet pairing correlations due to the on-site Coulomb repulsion, leading eventually to an anisotropic spin-singlet SC state~\cite{Hirsch85}. 

One year later, 
L. Caron and C. Bourbonnais~\cite{Bourbonnais86,Caron86} extended their theory for the generic \tm2x phase diagram to the metallic domain and proposed a gap equation with singlet superconductivity based on an interchain magnetic coupling with an attraction   deriving from an interchain exchange interaction overcoming the on-stack Coulomb repulsion. 
More recently, it has been recognized that, since the Cooper channel (responsible for superconductivity) and Peierls channel (responsible for spin/charge density wave orders) are both diverging at low temperature in 1D conductors, their behaviours in temperature should be treated  on equal footing. 
With the renormalization-group theory, one can take into account the interference between such diverging channels.
Such studies have been indeed  performed subsequently for Q1D conductors~\cite{Duprat01}.
An overview of the theory of 1D conductors can also be found in the textbook by T. Giamarchi \cite{Giamarchi04}. 

\begin{figure}[h]	
 \centerline{\includegraphics[width=1\hsize]{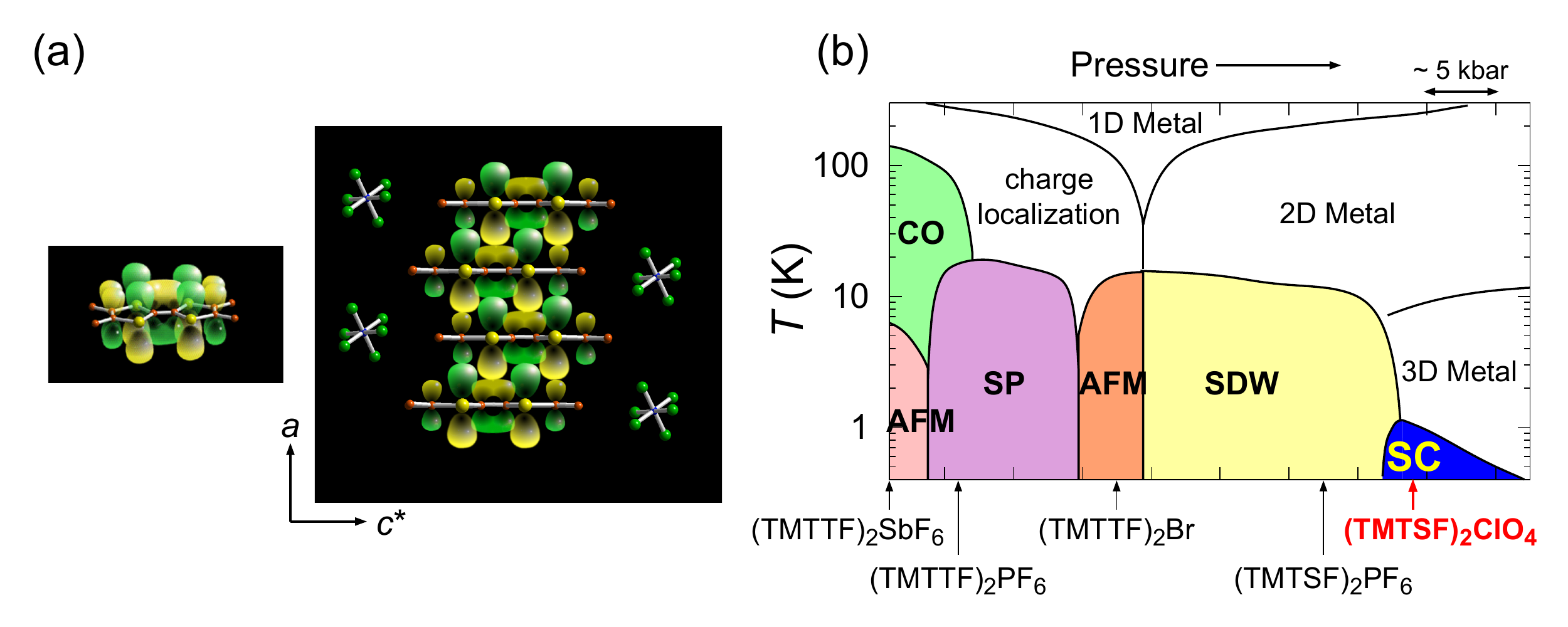}}
\caption{(a) Side view of the \tsm molecule (yellow and red dots are selenium and carbon  atoms respectively, hydrogens not shown) and \tmpns\ Q1D structure seen along the $b$ axis, \textit{courtesy} of J.Ch. Ricquier, IMN, Nantes. 
The yellow and green clouds around the atoms schematically present real-space distribution of molecular orbitals responsible for the electronic conduction.
 (b) Generic phase diagram for the  \tm2x family~\cite{Jerome91} based on experiments on the sulfur compound \tmttfsbfns.
The ambient pressure for this compound is taken as the origin for the pressure scale. 
The horizontal tics correspond to $\sim 5$~kbar interval. 
All colored phases are long-range ordered.
The curve between the 1D Metal and charge localization marks the onset of 1D charge localization, which ends around 15 kbar, slightly above \tmttfbrns.   
The 1D to 2D deconfinement occurs on the continuous curve in the higher-pressure regime. 
The curve between 2D and 3D regimes defines the upper limit for low-temperature 3D coherent domain. 
There exists a small pressure window around 45 kbar in this generic diagram where SC coexists with SDW according to Refs.~\cite{Vuletic02,Kang10,Narayanan14}.
\tmc is the only compound to exhibit superconductivity under ambient pressure.}
\label{molecule,structure,supra2} 
\end{figure}

For several experimental reasons, we are now entitled to attribute the pairing  in organic superconductivity to a mechanism which differs from the regular electron-phonon driven pairing in traditional superconductors.  First, superconductivity of Q1D Bechgaard salts shares a common border with magnetism as displayed on the generic diagram in Fig.~\ref{molecule,structure,supra2}(b). Second, strong antiferromagnetic  fluctuations exist in the normal state above \tc in the vicinity of the SDW phase, providing the dominant contribution to the nuclear hyperfine relaxation and also controlling the linear temperature dependence of electronic transport. Third, some experimental results point to the existence of a non-conventional pairing mechanism. These are summarized below.

\section{Basic properties of superconductivity}

\begin{figure}[h]	
 \centerline{\includegraphics[width=0.4\hsize]{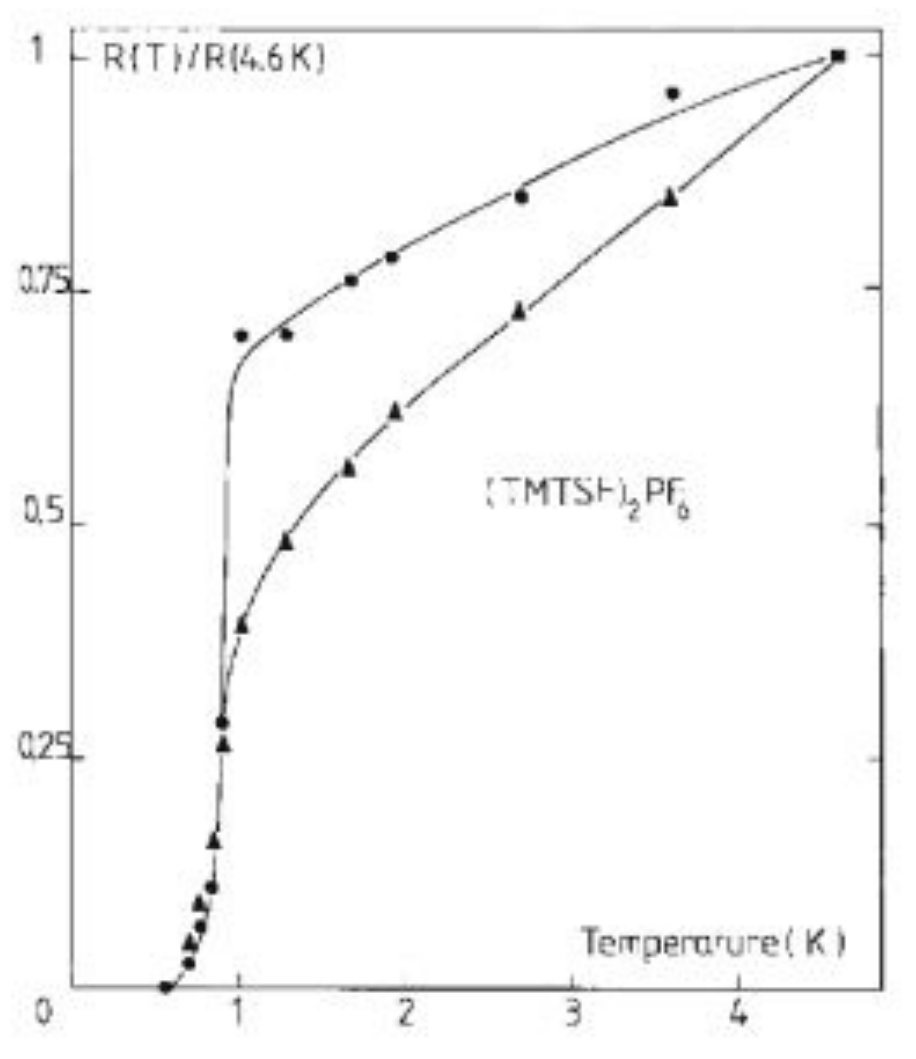}}
\caption{First observation of superconductivity in \tmpns\ under a pressure of 9 kbar~\cite{Jerome80}. The resistance of two samples is normalized to its value at 4.5~K.}
\label{PF6supra2.pdf} 
\end{figure}

\begin{figure}[htbp]			
\centerline{\includegraphics[width=0.4\hsize]{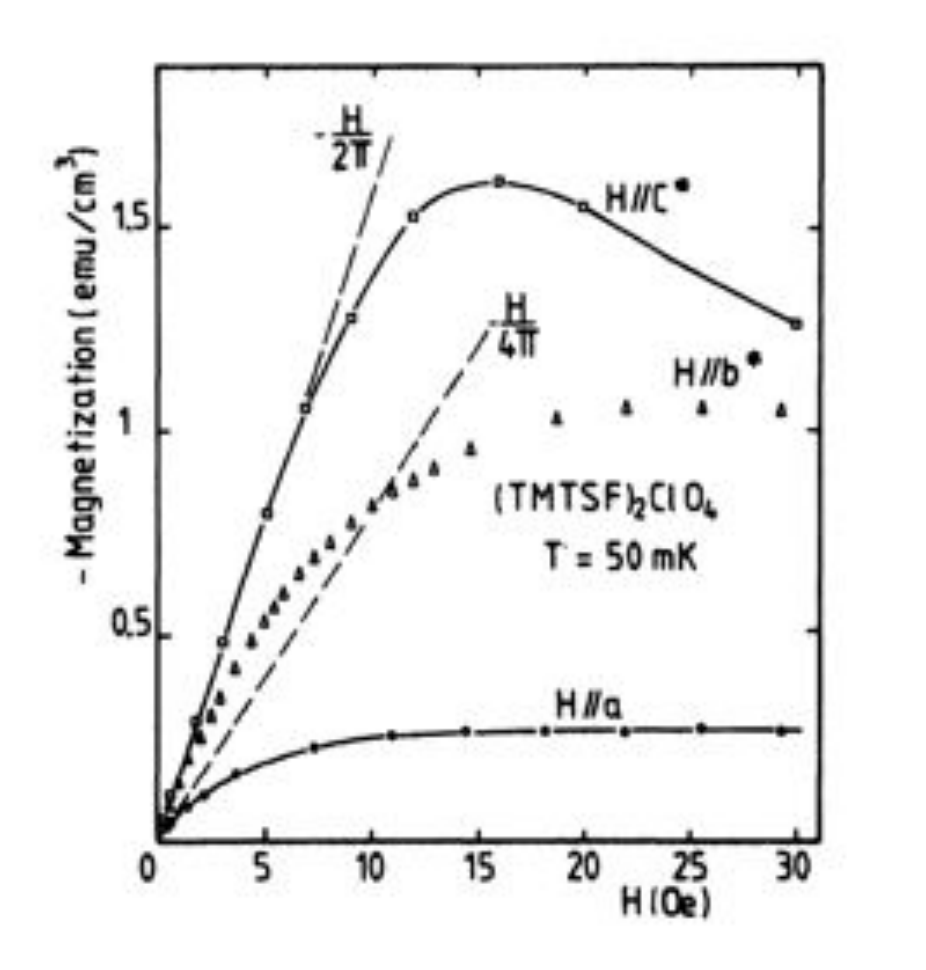}}
\caption{Diamagnetic shielding of \tmc at $T=0.05$~K  for magnetic fields oriented along the three crystallographic axes, from Ref.~\cite{Mailly82}.}
\label{MeissnerClO4.pdf}
\end{figure}

Although superconductivity in organic conductors  has first been  stabilized  under pressure~\cite{Jerome80} (see Fig.~\ref{PF6supra2.pdf}), more detailed investigations of  this phenomenon have been conducted in \tmc for experimental reasons since it is the only compound among the Q1D Bechgaard salt series that exhibits superconductivity (at 1.2~K) under ambient pressure.
Additional evidences for superconductivity in \tms2x
conductors came out from \tmc transport studies~\cite{Bechgaard81}, specific-heat measurements~\cite{Garoche1982.PhysRevLett.49.1346,Garoche1982.JPhysLett.43.L147} and Meissner flux expulsion (Fig.~\ref{MeissnerClO4.pdf})~\cite{Mailly82}. 
More recent specific-heat data are presented in Sec.~\ref{subsec:Specific_heat_data}.

Regarding evidences for the Meissner expulsion, the lower critical field $H\sub{c1}$ is obtained from the magnetization curves at low temperature. The obtained values are 0.2, 1, and 10 Oe along the $a$, $b\dash$, and $\cstar$ axes, respectively. 
Following the values for the upper critical fields $H\sub{c2}$ derived either from the Meissner experiments and the knowledge of the thermodynamical field~\cite{Mailly82} or from a direct measurements of transport, superconductivity is in the extreme type-II limit. The Ginzburg-Landau parameter $\kappa$ can even overcome 1000 when the field is along the $a$ axis due to the weak interchain coupling making the field penetration very easy for this external-field configuration.
An interpretation for the critical fields assuming the clean limit has been suggested in 1985~\cite{Gorkov85}. 
According to this theory, the slopes of $\Hcc(T)$ near $H=0$ should be given by:
\begin{align}
H_{{\mathrm c2}\,\parallel\,a} (T)& = \frac{98.7 \times 10^3}{t_{b\dash}  t_{\cstar}}\Tcz(\Tcz-T)\ \ ,
\label{eq:gl-a}\\
H_{{\mathrm c2}\,\parallel\,b\dash} (T)& = \frac{199 \times 10^3}{t_{\cstar}t_a}\Tcz(\Tcz-T)\ \ ,
\label{eq:gl-b}\\
H_{{\mathrm c2}\,\parallel\,\cstar} (T)&  = \frac{365 \times 10^3}{t_at_{b\dash}}\Tcz(\Tcz-T)\ \ ,
\label{eq:gl-c}
\end{align}
where $\Hcc$ is given in unit of kOe and the hopping integrals in K.
Derivation of the numerical factors are explained in Refs.~\cite{Gorkov85,Yonezawa2008.JPhysSocJpn.77.054712}.
This proposal was based on the microscopic expressions for the effective mass tensor in the Ginzburg-Landau equation near $\Tc$~\cite{Gorkov64}.

Given $\Hcc$ slopes near \tc of \tmc from transport studies~\cite{Yonezawa2008.PhysRevLett.100.117002,Yonezawa2008.JPhysSocJpn.77.054712}, $dH_{\mathrm{c}2\,\parallel\,a}/dT = -67$~kOe/K, $dH_{\mathrm{c}2\,\parallel\,{b\dash}}/dT =-36$~kOe/K, and $dH_{\mathrm{c}2\,\parallel\,\cstar}/dT=-1.5$~kOe/K, Eqs.~(\ref{eq:gl-a}-\ref{eq:gl-c}) lead to band parameters $t_a : t_{b\dash} : t_{\cstar}$ = 1200, 310, and 7~K, respectively.
If one uses slopes from a thermodynamic study~\cite{Yonezawa2012.PhysRevB.85.140502R}, $dH_{\mathrm{c}2 \,\parallel\,a}/dT = -81$~kOe/K, $dH_{\mathrm{c}2\,\parallel\,b\dash}/dT =-23$~kOe/K, and $dH_{\mathrm{c}2 \,\parallel\,\cstar}/dT=-1.1$~kOe/K, we obtain $t_a : t_{b\dash} : t_{\cstar}$ = 1800, 250, and 6~K.
These values are in reasonable agreement with the realistic band parameters~\cite{Ishiguro98}

\begin{figure}[h]			
\centerline{\includegraphics[width=0.5\hsize]{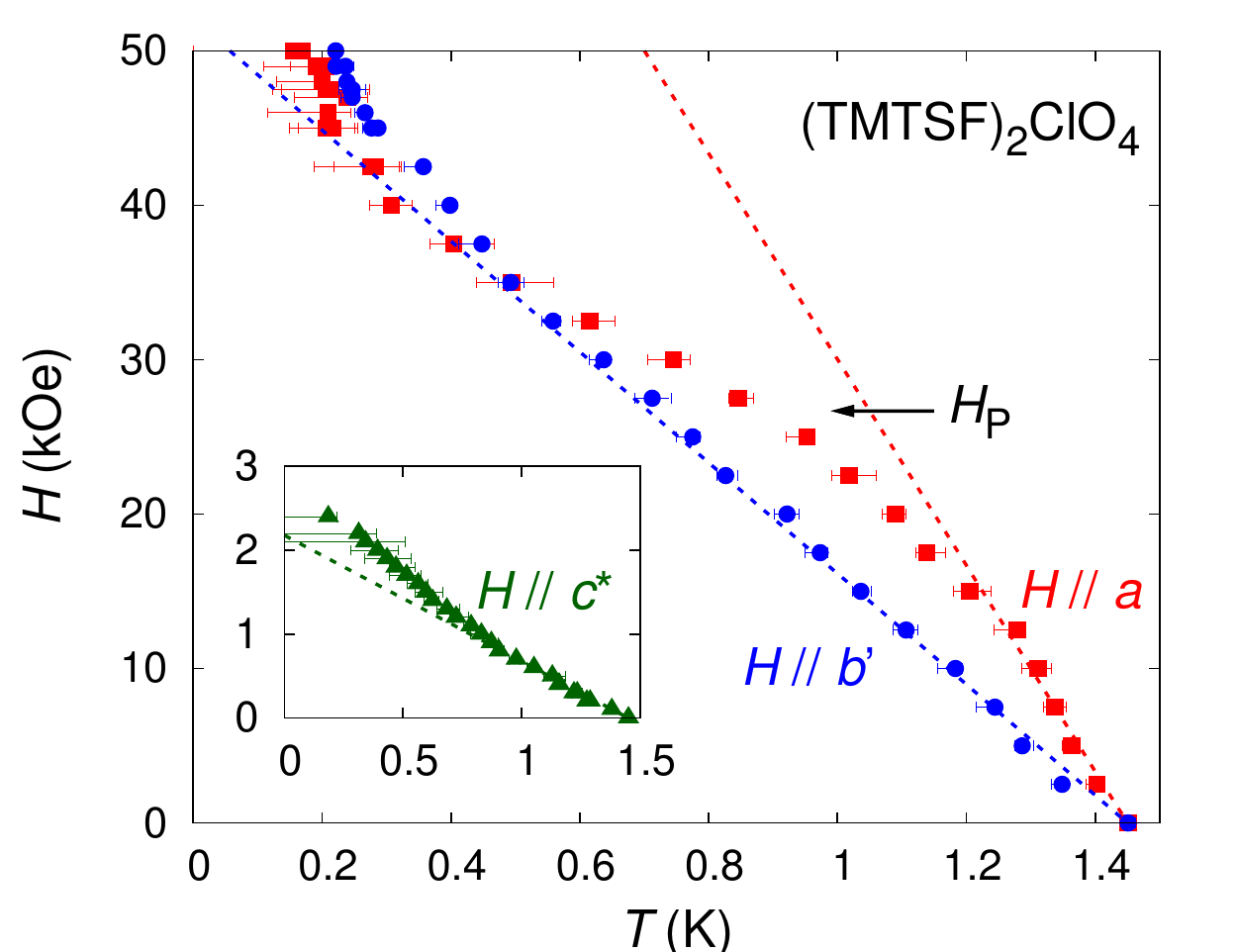}}
\caption{Critical fields of \tmc determined from the onset temperature of the $\cstar$-axis resistance $\Tco$ for fields along the three principal axes with an indication for the Pauli limit at low temperature. 
The figure is taken from Ref.~\cite{Yonezawa2008.PhysRevLett.100.117002}.
\label{fig:phase-diagram}}
\end{figure}

From the slopes of $\Hcc(T)$, one can also deduce the SC coherence lengths $\xi_i$ ($i = a, b\dash, \cstar$), by using formulae $\Hcc\sps{orb} = -0.73 \Tcz\, d\Hcc(T)/dT|_{T=\Tcz}$~\cite{Helfand1966PhysRev} and $H\sps{orb}_{\mathrm{c}2\,\parallel\,i} = \varPhi_0/(2\pi \xi_j\xi_k)$, where $\varPhi_0$ is the flux quantum. We obtain $(\xi_a, \xi_{b\dash}, \xi_{\cstar}) = (620~\mathrm{\AA}, 330~\mathrm{\AA}, 14~\mathrm{\AA})$ from the transport phase diagram~\cite{Yonezawa2008.JPhysSocJpn.77.054712}\footnote{The coherence length values in Ref.~\cite{Yonezawa2008.JPhysSocJpn.77.054712} should be multiplied by $\sim 1.4$ because of a trivial calculation error.}, and $(\xi_a, \xi_{b\dash}, \xi_{\cstar}) = (1100~\mathrm{\AA}, 300~\mathrm{\AA}, 14~\mathrm{\AA})$ from thermodynamic phase diagram.
The obtained coherence lengths are quite anisotropic reflecting the Q1D nature of \tmcns.
Note that the coherence lengths $\xi_a$ and $\xi_{\cstar}$ from the transport may be underestimated because of the enhancement of $\Hcc \parallel b\dash$ due to the field-induced dimensional crossover~\cite{Lebed1986.JETPLett.44.114,Zhang2007AdvPhys}.
Also notice that the coherence lengths are much shorter than the mean free path along the $a$ axis, $l_a \sim 1.6~\muup$m~\cite{Yonezawa2008.JPhysSocJpn.77.054712}.
Thus, this system is well within the clean limit $\xi \ll l$.

\section{Non-s-wave superconducting nature in \tms2x}\label{sec:non-s-sc}

In general, SC states can be classified by symmetries associated with the SC wave function.
The SC wave function should be odd under commutation of electrons, because electrons are Fermions.
The most simple state, assumed in the original BCS theory and indeed realized in most of superconductors, is the spin-singlet  state (i.e. spin state being represented as $\propto |\!\uparrow\downarrow - \downarrow\uparrow\rangle/\sqrt{2}$, with the total spin $S = 0$) with a $k$-independent isotropic gap. 
Such a state is called the $s$-wave SC state, in analogy to the atomic $s$ orbitals, which is isotropic in the real space.
However, gaps anisotropic in the $k$ space are possible, e.g. when magnetic interactions are responsible for the Cooper pairing.
Such gaps can be classified as $p$, $d$, $f$, \ldots waves, depending on the rotational symmetry breaking associated with the wave function,  again analogously to the atomic wave function. 
For odd-parity states such as $p$, $f$, \ldots wave states, odd-commutation condition require the spin state to be of spin-triplet nature (total spin $S=1$, combinations of spin states $|\!\uparrow\uparrow\rangle$, $|\!\uparrow\downarrow + \downarrow\uparrow\rangle/\sqrt{2}$, $|\!\downarrow\downarrow\rangle$).
Notice that the spin part for a triplet state is even under commutation.
Investigation of non-$s$-wave pairing has been one of the central topics of condensed-matter physics for more than 30 years.

Since the SC phase of \tmc is located next to the spin-density wave (SDW) phase as shown in Fig.~\ref{molecule,structure,supra2}(b), possibility of non-$s$-wave pairing mediated by spin fluctuation has been proposed.
Experimentally, early evidences for non-$s$-wave pairing in \tmc have been obtained with the spin-lattice relaxation rate $1/T_1$ measurement with the nuclear magnetic resonance (NMR) technique, which revealed absence of the coherence peak just below $\Tc$ as well as the power-law behavior at lower temperatures~\cite{Takigawa1987}.
This behavior is theoretically interpreted as a consequence of non-$s$-wave pairing~\cite{Hasegawa1987.JPhysSocJpn.56.877}.
Strong suppression of superconductivity by non-magnetic impurities was revealed by using alloyed samples (TMTSF)\sub{2}(ClO\sub{4})$_{1-x}$(ReO\sub{4})$_{x}$~\cite{Joo2004,Joo2005}, as described in detail in Sec.~\ref{sec:non-magnetic-defects}.
Observation of a $\sqrt{H}$ dependence of the low-temperature specific heat~\cite{Yonezawa2012.PhysRevB.85.140502R}, as well as the temperature dependence of the specific heat in zero field described below, also provide strong evidence for nodal SC state.
Furthermore, the in-plane field-angle dependence of the specific heat provides information on the location of nodes, as explained in Sec.~\ref{sec:Magneto-calorimetric}.
We note that several experiments claim fully gapped states: the thermal conductivity~\cite{Belin1997} reveals electronic thermal conductivity vanishes exponentially below $\Tc$ after subtraction of phonon contribution; in-field muon spin rotation ($\muup$SR)~\cite{Pratt2013.PhysRevLett.110.107005} revealed the temperature dependence of the penetration depth suggesting a fully gapped states but only in magnetic fields. 
Nevertheless, we believe that so far nodal SC scenario has been accumulating more direct evidence.
We also note that zero-field $\muup$SR measurement~\cite{Pratt2013.PhysRevLett.110.107005} could not detect spontaneous time-reversal symmetry breaking (i.e. spontaneous magnetization) in the SC state, excluding possibility of ``chiral'' SC state.
Experiments on superconductivity in \tms2x are also reviewed excellently in Refs.~\cite{Lee2006.JPhysSocJpn.75.051011.Review,Lebed2008.Book,Jerome2012.JSupercondNovMagn,Brown2015.PhysicaC.514.279}.

Theoretically, as already explained, spin-fluctuation pairing mechanism in \tms2x has been proposed as early as 1986~\cite{Emery86}.
A tremendous amount of theories have been proposed, because of the simplicity of the Q1D electronic structure in \tms2x, as well as stimulation by interesting experiments. 
Microscopic theories considering spin and/or charge fluctuations have proposed unconventional SC state, not only spin-singlet $d$-wave-like states, but also spin-triplet $p$-wave-like or $f$-wave-like states, based on methods such as
random phase approximation (RPA) or fluctuation exchange (FLEX) theories~\cite{%
  Shimahara1989.JPhysSocJpn.58.1735,%
  Kino1999.JLowTempPhys.117.317,%
  Kuroki2001.PhysRevB.63.094509,%
  Tanaka2004,%
  Aizawa2009.PhysRevLett.102.016403,%
  Aizawa2009.JPhysSocJpn.78.124711,%
  Mizuno2011.PhysicaC.471.49},
quantum Monte Carlo method~\cite{%
  Kuroki1999.PhysRevB.60.3060,%
  Kuroki2004.PhysRevB.69.214511,%
  Kuroki2005.JPhysSocJpn.74.1694},
perturbation theory~\cite{%
  Nomura2001.JPhysSocJpn.70.2694},
and RG theory~\cite{%
  Fuseya2005.JPhysSocJpn.74.1263,%
  Nickel05,%
  Nickel06,%
  Bourbonnais2009.PhysRevB.80.085105,%
  Cho2013.PhysRevB.88.064505%
}.
Considering only the one pair of Fermi surface sheets, which is the electronic band structure for \tmp6, $s$ or $p$-wave-like states can be fully gapped, whereas $d$ or $f$-wave-like states should have nodes on the Fermi surface.
In case of \tmcns, the Fermi surface consists of two pairs of sheets at low temperatures because of the band folding due to the anion ordering below $T\sub{AO} = 24$~K~\cite{Pevelen2001EurPhysJB}.
For such ``folded'' Fermi surfaces, it has been pointed out that a fully-gapped $d$-wave-like state is also possible~\cite{Shimahara2000.PhysRevB.61.R14936}.
For more details of theories, see review articles such as Refs.~\cite{Kuroki2006.JPhysSocJpn.75.051013,Zhang2007AdvPhys,Lebed2008.Book}

\subsection{Specific heat data}
\label{subsec:Specific_heat_data}

\begin{figure}[htb]
\begin{center}
\includegraphics[width=0.8\hsize]{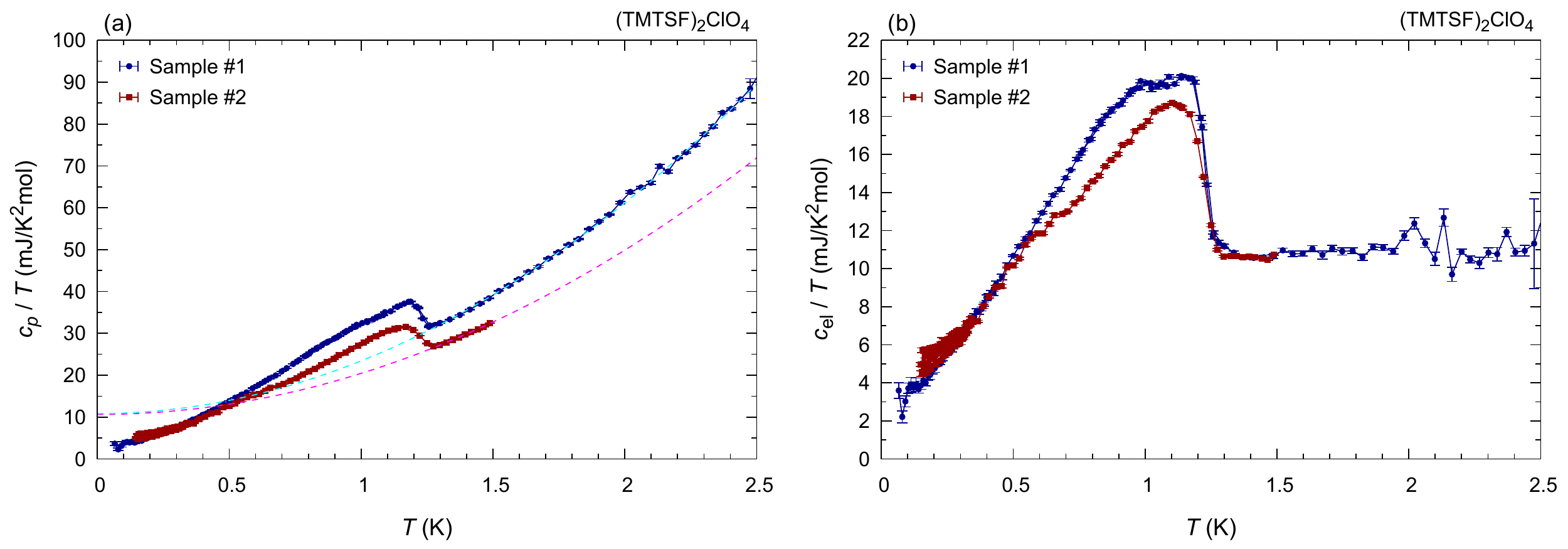}
\caption{(a) Temperature dependence of the specific heat of \tmc.
We present data for two different single crystals, Sample~\#1 (0.257~mg; blue circles) and Sample~\#2 (0.364~mg; red squares).
The broken curves are fitting results with the Sommerfeld-Debye formula $C_p/T=\gamma\sub{e} + \beta\sub{p}T^2$ to the normal state data ($T>1.3$~K). 
Resulting fitting parameters are $\gamma\sub{e} = 10.8 \pm 0.2 $~mJ/K\sps{2}mol and $\beta\sub{p}= 12.6 \pm 0.1$~mJ/K\sps{4}mol for Sample~\#1, and $\gamma\sub{e} = 10.6 \pm 0.4 $~mJ/K\sps{2}mol and $\beta\sub{p}= 9.8 \pm 0.2$~mJ/K\sps{4}mol for Sample~\#2.
(b) Electronic specific heat $C\sub{el}/T$ of the two samples.
 \label{fig:C-T}
}
\end{center}
\end{figure}

Recent new results of the temperature dependence of the specific heat $C_p$ are presented in Fig.~\ref{fig:C-T}.
This data is obtained by the ac technique~\cite{Sullivan1968} and using only one single crystal.
To improve the accuracy of the obtained data, we measured the dependence of the temperature oscillation amplitude $T\sub{ac}$ on the heater-current frequency $f$ and we fitted the $T\sub{ac}(f)$ data with the theoretical function $T\sub{ac}(f)= P/(8\pi fC_p)[1 + (4\pi f \tau_1)^2 + (4 \pi f \tau_2)^{-2}]^{-1/2}$ to obtain $C_p$, where $\tau_1$ and $\tau_2$ are the external and internal relaxation rates, respectively.
More details will be published elsewhere.
In Fig.~\ref{fig:C-T}, we compare results for different samples.
Both samples exhibits sharp anomaly at around $\Tc \sim 1.2$~K, indicating bulk superconductivity.
The electronic specific-heat coefficient is found to be  $\gamma\sub{e}=10.6$--10.8~mJ/K\sps{2}mol, in good agreement with the previous works ($\gamma\sub{e}=10.5$~mJ/K\sps{2}mol)~\cite{Garoche1982.PhysRevLett.49.1346,Garoche1982.JPhysLett.43.L147}, although the phononic specific heat coefficient exhibits $\sim 20$\% variation depending on samples, but still comparable to $\beta\sub{p} = 11.4$~mJ/K\sps{4}mol reported in Ref.~\cite{Garoche1982.JPhysLett.43.L147}. 
In addition, it can be checked from the data in Fig.~\ref{fig:C-T}(b) that the entropy of the SCstate at \tc equals that of the normal state at the same temperature within $\sim 13$\% for both samples. 

Figure~\ref{fig:C-T}(b) displays several features supporting a non-$s$-wave pairing state.
Firstly, the height of the specific-heat jump at $\Tc$, $\varDelta C$, nearly equals to $\gamma\sub{e}\Tc$. 
This is notably smaller than the expectation of the BCS theory, in which $\varDelta C/\gamma\sub{e}\Tc$ is expected to be 1.43.
Instead, it is known that $\varDelta C/\gamma\sub{e}\Tc$ can be smaller than 1.43 if the SC gap has substantial anisotropy.
In particular, if the gap has line nodes the ratio can be even smaller than 1.0~\cite{HasselbachK1993.PhysRevB.47.509,Emery83}.
Secondly, $C\sub{el}/T$ exhibits linear temperature dependence in a wide temperature range below $\sim 0.7$~K.
Such linear behavior also evidences nodal SC gap.
The finite intercept  for the linear extrapolation of $C\sub{el}/T$  to zero temperature is expected in case of a finite  elastic scattering time~\cite{Sun95}. Using the data for sample \# 1, the residual density of states   amounts to about 18.5\% the value of the normal state  according to Fig.~\ref{fig:C-T}(b). This will  be further commented on in the next  section.

\subsection{Non magnetic defects}
\label{sec:non-magnetic-defects}

A basic property of the \textit{s}-wave superconductivity proposed in the BCS theory is the isotropic ($k$-independent) gapping on the Fermi surface.  
Hence, no pair breaking is expected from the scattering of electrons against spinless impurities~\cite{Anderson59}, since such scatterings essentially just mixes and averages gaps at different $k$ positions. 
Experimentally, this property has been verified in non-magnetic dilute alloys of \textit{s}-wave superconductors and
provided a strong support to the BCS model of conventional \textit{s}-wave superconductors. 
However, the condition for isotropic gap is no longer met for the case of non-$s$-wave pairing, in which the average of the gap $\varDelta(\bm{k})$ over the Fermi surface vanishes due to sign changes in $\varDelta(\bm{k})$, i.e. $\sum_{\text{FS}}\varDelta(\bm{k}) \sim 0$. 
Consequently, $\Tc$ for these superconductors should be strongly affected by any non-magnetic scattering, cancelling out positive and negative parts of the gap. 
Theories on effects of non-magnetic impurities on $\Tc$ in such superconductors have been deduced by generalizing conventional pair-breaking theory for magnetic impurities in $s$-wave superconductors.
Then the famous relation,
\begin{align}
\ln\left(\frac{\Tc^0}{\Tc}\right)=\psi\left(\frac{1}{2}+\frac{\alpha \Tc^0}{2\pi \Tc}\right)-\psi\left(\frac{1}{2}\right),
\label{eq:Digamma}
\end{align}
is obtained~\cite{Maki04,Larkin65}, with $\psi(x)$ being the Digamma function and $\alpha = \hbar /2 \tau k_{B}\Tc^0$ the depairing parameter related to the elastic scattering time $\tau$.
Experimentally, it has been found that this relation holds for non-$s$-wave superconductors such as Sr\sub{2}RuO\sub{4} ($\Tc = 1.5$~K; most likely a $p$-wave spin-triplet superconductor)~\cite{Mackenzie2003RMP,Maeno2012.JPhysSocJpn.81.011009}.

It is also the remarkable sensitivity  of organic
superconductivity to irradiation detected in the early years~\cite{Bouffard82,Choi82} that led Abrikosov to suggest the possibility of triplet pairing in these materials~\cite{Abrikosov83}.
A more recent investigation of the influence of non magnetic defects on organic superconductivity has been conducted following a procedure which rules out the addition of possible magnetic impurities, which is the case for X-ray irradiated samples~\cite{Miljak80}. Attempts to synthesize non-stoichiometric compounds have not been successful for these organic salts. However, what turned out to be feasible is an iso-electronic  anion solid solution keeping the charge transfer constant. One attempt has been to  create non-magnetic  disorder through the synthesis of solid solutions with centrosymetrical anions such as \as\ and \sb. This attempt turned out to be unsuccessful as the  effect of disorder happened to be very limited with only a minute effect on \tc~\cite{Traettebergthesis}. 

\begin{figure}[h]	
\centerline{ \includegraphics[width=0.5\hsize]{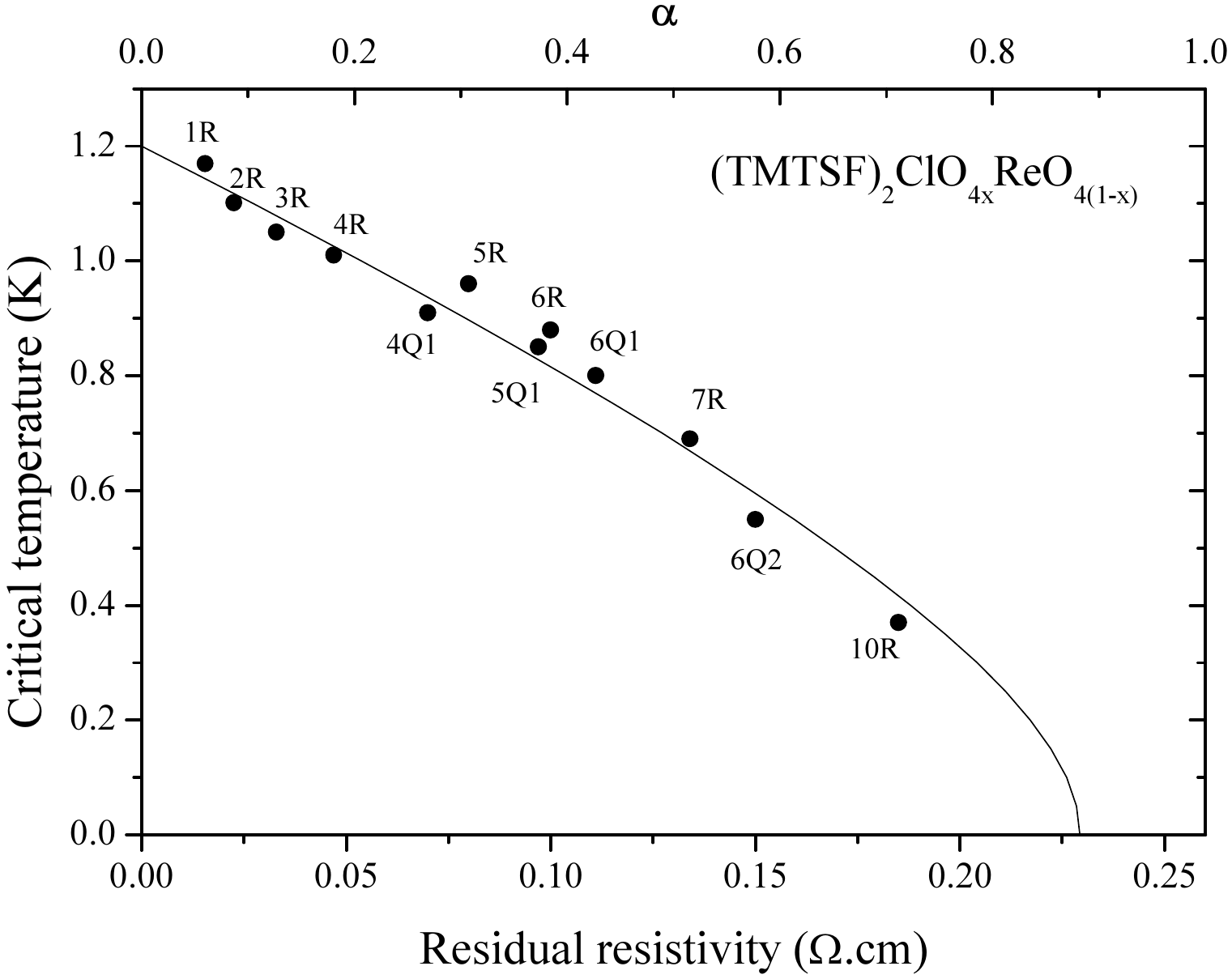}}
\caption{Phase diagram of \tmxns, governed by non magnetic disorder.
The data are obtained by newly analysing the temperature dependence of resistivity reported in Ref.~\cite{Joo2005} (see text). Points with labels ``R'' refer to  very slowly cooled samples in the
R-state (the so-called relaxed state) with different \R\ contents, whereas points with labels ``Q'' refer to quickly cooled samples in the quenched state.
A sample with $\rho_0=0.27~\Omega \mathrm{cm}$ {\it i.e}, beyond the critical defect concentration, is metallic down to the lowest temperature of the experiment. The continuous curve is a fit of Eq.~\eqref{eq:Digamma} to the data with $\Tc{^0}=1.23$~K.}
\label{TcvsRho0.pdf} 
\end{figure}

Another scheme with which non-magnetic defects can be introduced in a controlled way for non-centro-symetrical anions in the \tms2x series is either by fast cooling preventing the complete ordering of the tetrahedral \cl\ anions or by introducing ReO\sub{4} anions to the ClO\sub{4} site by making the solid solution \tmxns.
As displayed on Fig.~\ref{TcvsRho0.pdf}, superconductivity in the solid solution is  suppressed  and the reduction in $\Tc$ is clearly related to the residual resistivity, the enhancement of the elastic scattering in the normal state. 
The data on Fig.\ref{TcvsRho0.pdf} show that the relation \tc versus $\rho_0$ follows Eq.~\eqref{eq:Digamma} with good accuracy with $\Tc{^0}=1.23$~K. 

At this stage, it is worth pointing out that the determination of the residual resistivity is not a trivial matter.  Various procedures have been used in the literature. First, the resistivity displays an usual quadratic temperature dependence both above the anion ordering temperature $T\sub{AO}= 24$~K and below down to approximately 10~K. Consequently, a first attempt to determine  $\rho_0$ was to extrapolate $\rho(T)$ down to zero temperature the quadratic behaviour observed between $T\sub{AO}$ and 10~K. It turned out that $\rho_0$ is rather ill defined with this procedure (see Ref.~\cite{Joo2005}). Second, another procedure was to use a linear extrapolation of the temperature dependence below 10~K down to $\Tc$, leading to lower values of $\rho_0$~\cite{Joo2004}. 
However, several recent re-analysis of the temperature dependence of the resistivity in the neighborhood of \tc in pure \tmcns~\cite{DoironLeyraud2010.EurPhysJB.78.23} and in the alloy series \cite{Auban-Senzier2011.JPhysCondensMatter.23.345702} have emphasized the existence of two different regimes: a regime between 10 and 2~K where the single particle scattering is dominated by antiferromagnetic fluctuations leading in turn to a linear dependence, and another regime between 2~K and \tc where the downturn of the resistivity can be ascribed to the sliding of SDW waves without any transverse coherence in the vicinity of an antiferromagnetic order. The latter is not accessible in \tmc since it would require  a negative pressure as shown by elongation experiments along the $b\dash$ axis~\cite{Kawada2007.JPhysSocJpn.76.114710}. 
The procedure used to derive $\rho_0$ in Fig.~\ref{TcvsRho0.pdf} is a linear extrapolation to zero temperature  of the linear regime between 2 and 10~K dominated by scattering against AF fluctuations.
This procedure should be rather accurate, in particular, in \tmcns.

It has been checked that the additional scattering cannot be ascribed to magnetic scattering with the electron paramagnetic resonance (EPR) technique, which shows no additional traces of localized spins in the solid solution.
Thus, the data in Fig.~\ref{TcvsRho0.pdf} 
 cannot be reconciled with the picture of a SC gap keeping a constant sign over the whole
$(\pm k_F)$ Fermi surface. They require a picture of pair breaking in a superconductor with an anisotropic gap
symmetry.

It is interesting to compare the residual density of states predicted by theories with experimental data. 
Figure~\ref{TcvsRho0.pdf} shows that the depairing  parameter of the pristine sample amounts to about 6.25\%  the critical value for the suppression of superconductivity. Given the ratio $\Gamma/\Gamma_0=0.0625$ for the pristine sample where $\Gamma$ is the scattering rate, the calculation of Sun and Maki~\cite{Sun95} leads in turn to a residual density of states $N(0)$= 0.26$N_0$ which is fairly close to the residual density of states derived from our specific heat experiments, see the previous subsection. In the NMR data in Ref.~\cite{Shinagawa2007}, the spin lattice relaxation rate $1/T_1$ below 0.2 K amounts $\sim 25$\% of that in the normal state. Since, $1/T_1$ is proportional to the square of the density of states, the observed residual value of $1/T_1$ corresponds to $\sim 50$\% of the density of states remaining in the SC state. 
Such a residual density of states from NMR compares very favorably with the value 18.5\% provided by the measurement of the electronic specific heat. The larger value found by NMR can be attributed to the field dependence of the density of states, as reported in Ref.~\cite{Yonezawa2012.PhysRevB.85.140502R}, since NMR data have been taken under magnetic field $\mu_0H=0.96$~T along the $b'$ axis or 1.3~T along the $a$ axis.
  
The influence of non-magnetic impurities on the SC phase implies the existence of  positive as well as negative values for the SC order parameter on the Fermi surface. It precludes the usual case of \textit{s}-symmetry but is still unable to 
 discriminate between  two possible options namely,  singlet-\textit{d (g)} or triplet-\textit{p (f)}~\cite{Nickel05} (see Fig.~\ref{Nodes.pdf}).

\subsection{Spin susceptibility in the superconducting phase}
\label{sec:spin-susceptibility}

The detailed study of the behaviour of static and dynamic properties electron spins has been undertaken via the $^{77}\mathrm{Se}$ Knight shift and $1/T_{1}$ measurements across \tc in the compound \tmcns~\cite{Shinagawa2007}.

\begin{figure}[htbp]	
 \centerline{\includegraphics[width=1\hsize]{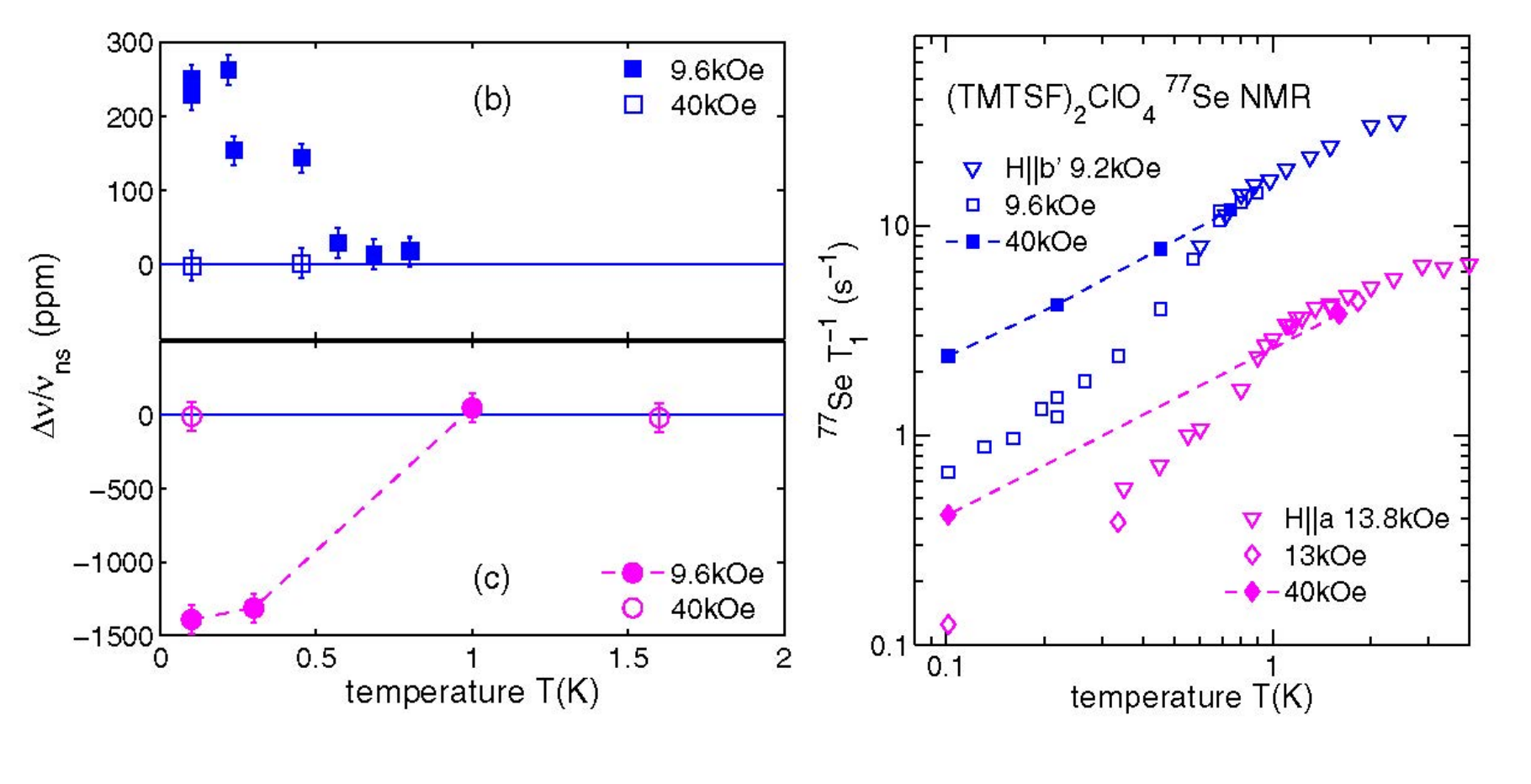}}
\caption {$^{77}$Se Knight shift (a) and $1/T_1$ vs $T$ (b) for \tmcns, for $H// b'$ and $a$, according to reference ~\cite{Shinagawa2007}. The sign of the variation  of  Knight shift at \tc depends on the sign of the hyperfine field. 
A linear temperature dependence of the relaxation rate is recovered at very low temperature signaling the existence of unpaired carriers at the Fermi level.}
\label{KsandT1} 
\end{figure}

\begin{figure}[htbp]	
 \centerline{\includegraphics[width=0.8\hsize]{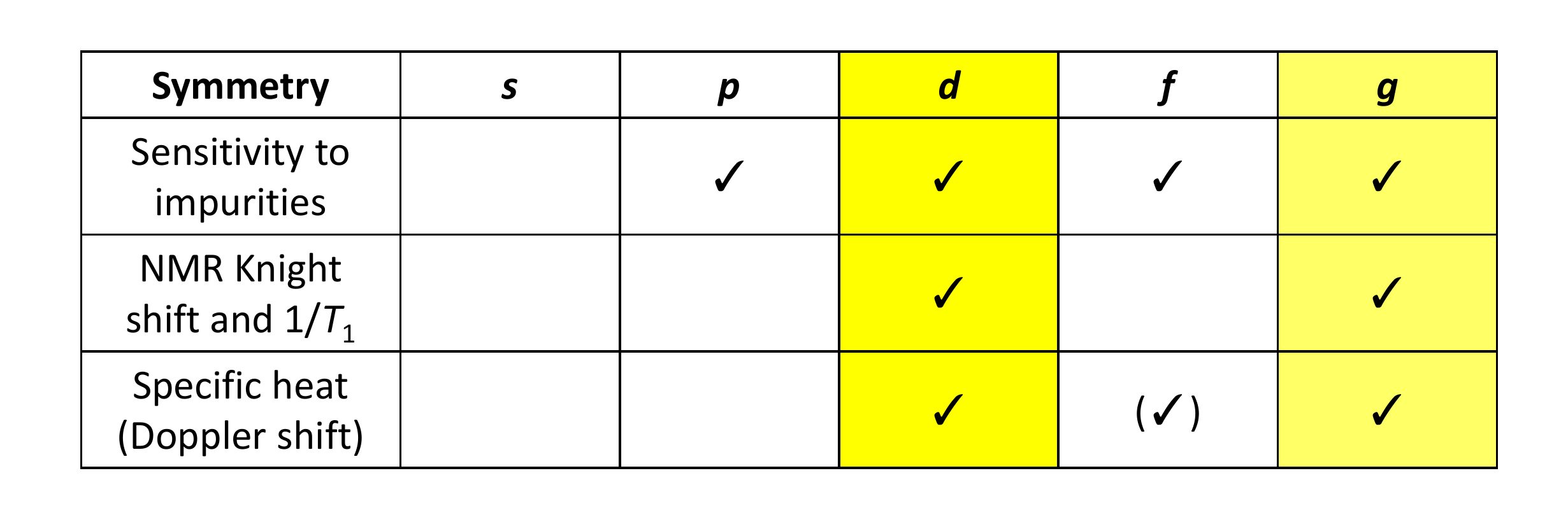}}
\caption{Possible gap symmetries agreeing with the different  experimental results. 
The spin-singlet \textit{d}-wave  (or \textit{g}-wave) symmetry is the only symmetry agreeing with all experiments (yellow columns on line).}
\label{Nodes.pdf} 
\end{figure}

The Knight shift is revealed to decrease below $\Tc$, as presented in Fig.~\ref{KsandT1}(a), providing solid evidence in favour of spin-singlet pairing. Furthermore, the temperature dependence of the relaxation rate shown in Fig.~\ref{KsandT1}(b) does not display the exponential behaviour  expected in a regular fully gapped $s$-wave superconductor but instead a power-law dependence below \tc with a linear regime establishing below 0.2~K showing that there is a non-zero density of states at the Fermi level. Because $1/T_1T$ is proportional to the square of the density of states, the observed residual $1/T_1T$ value amounting 25-30\% of the density in the normal state indicates that the density of states is recovered by 50\% at $\mu_0H = 0.96$~T for $H\parallel b\dash$ and $\mu_0H = 1.3$--1.4~T for $H\parallel a$.

Moreover, a steep increase of the spin-lattice relaxation rate versus magnetic field  for both field orientations parallel to $a$ and $b\dash$ has provided  the evidence for a sharp cross-over or even a phase transition  occurring at low temperature under magnetic field  between 1 and 2 Tesla from  the low field \textit{d}-wave singlet phase and a high field regime exceeding the paramagnetic limit $H\sub{P}$ being either a triplet-paired state~\cite{Shimahara00,Belmechri07} or an inhomogeneous Fulde-Ferrell-Larkin-Ovchinnikov state~\cite{Fulde1964,Larkin1965}.
The nature of this high-field phase is further discussed in Sec.~\ref{sec:high-field-state}.

According to the NMR data,  a spin-singlet pairing is clearly established for the SC state of \tmcns. Quite a different  situation had been claimed in an earlier $^{77}\mathrm{Se}$ NMR  study of the SC phase of \tmpns~\cite{Lee01}. The $^{77}\mathrm{Se}$ Knight shift revealing no change through \tc had been taken as a strong suggestion in favor of spin-triplet superconductivity in \tmpns.  
However, it can be noticed that the experiment
in \tmpns\ had been conducted under a relatively high magnetic field of $\mu_0 H= 1.43$~T aligned along the most conducting   $a$ axis. 
Following the results in \tmc and, in particular, the relaxation-rate data of the latter compound displayed in Figure~\ref{fig:Shinagawa2007_T1-H}, a field of $1.43$~T may have been high enough to place the sample in to the high-field SC phase discussed in Sec.~\ref{sec:high-field-state}, for which the density of states does not reveal any noticeable change through the SC transition. This is corroborated by the data of the Korringa relaxation rate  in \tmpns, showing hardly any change after crossing \tc (see Fig.~5 in Ref.~\cite {Lee01}). In conclusion, although different  symmetry for the order parameter in \tmc and \tmpns\ cannot be totally ruled  out, we consider such a scenario as quite unlikely.   

\subsection{Magneto-calorimetric studies}
\label{sec:Magneto-calorimetric}

In addition to the spin state, the orbital gap structure of the SC state is also fundamentally important information.
As we explained in the preceding sections, evidence for non-s-wave pairing state in \tms2x\ had been accumulated.
To reveal more precise gap structures, one of the common techniques is to measure the field-angle-dependent quasiparticle excitations.
As first proposed by Volovik~\cite{Volovik1993.JETPLett.58.469}, superconductors with nodes (or zeros) exhibit field dependent quasiparticle excitation with momentum close to the nodal position, induced by the energy shift $\delta\omega$ caused by the supercurrent surrounding vortices penetrating the sample.
Such field-induced excitations are now called the Volovik effect.
This quasiparticle excitation is also field-direction dependent~\cite{Vekhter1999}, because $\delta\omega$ is proportional to the inner product of the Fermi velocity $\vF$ at the node and the superfluid velocity $\vs$, the latter being in turn  perpendicular to the applied field.
Thus, if one rotates the magnetic field within a certain plane, it is expected that the quasiparticle density of states oscillates as a function of the field angle.
Such oscillation can be detected by measuring, for example the specific heat or thermal conductivity while rotating the magnetic field within the conducting plane.
Indeed, such studies have been widely performed in three-dimensional (3D) or quasi-two-dimensional (Q2D) tetragonal systems such as CeCoIn\sub{5}~\cite{Izawa2001.PhysRevLett.87.057002,An2010.PhysRevLett.104.037002}, YNi\sub{2}B\sub{2}C~\cite{Izawa2002.PhysRevLett.89.137006}, 
Sr\sub{2}RuO\sub{4}~\cite{Deguchi2004.PhysRevLett.92.047002,Deguchi2004.JPhysSocJpn.73.1313}, and many other materials~\cite{Sakakibara2007.JPhysSocJpn.76.051004.review}.

\begin{figure}[htb]
\begin{center}
\includegraphics[width=0.9\hsize]{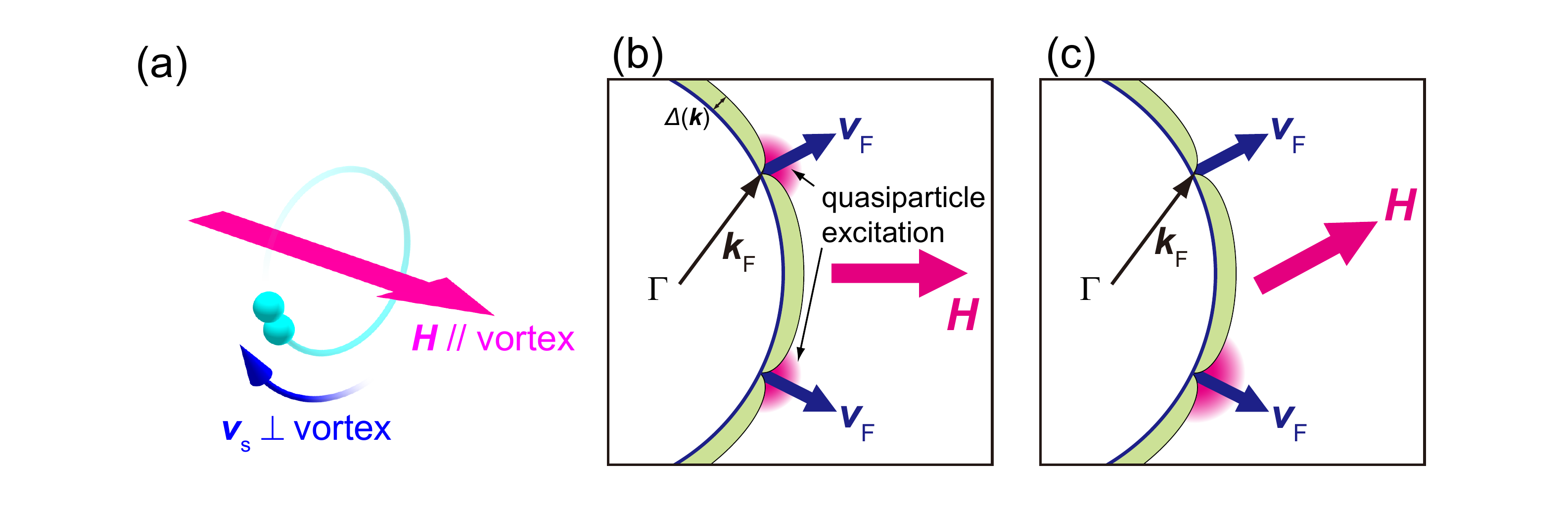}
\caption{Schematic description of the Volovik effect in a superconductor with gap nodes or zeros. (a) Supercurrent flowing around magnetic vortices. Supercurrent velocity $\vs$ is perpendicular to the vortex direction, namely the magnetic field direction. (b) Quasiparticle excitation around gap nodes excited by the Volovik effect. (c) Quasiparticle excitation when the field is parallel to the Fermi velocity at a node. In such situation, the excitation at this node is zero, since $\vs\cdot\vF = 0$ at this node. \label{fig:Volovik-effect}
}
\end{center}
\end{figure}

In contrast to Q2D systems, the story for Q1D systems is not so simple, because substantial in-plane anisotropy of $\Hcc$ leads to pronounced specific-heat oscillation as a function of the in-plane field angle, even concealing the oscillation originating from the gap anisotropy.
In addition, one should be careful that the Fermi velocity $\vF$ and the Fermi wavenumber $\kF$ are {\em not} necessarily parallel to each other (see Fig.~\ref{fig:Volovik-effect}(b)). 
More specifically, $\vF$, which is parallel to the gradient of the quasiparticle energy $\varepsilon(\bm{k})$ in the reciprocal space and thus is perpendicular to the Fermi surface, is not always parallel to $\kF$, which is the vector pointing at a $k$ position on the Fermi surface from the origin of the reciprocal space (the $\Gamma$ point).
In Q1D systems, this is almost always the case.
Therefore, even if the specific-heat oscillation originating from the gap is observed, the field direction where the specific heat exhibits an anomaly does not have any direct relation to the gap node position in the $k$ space:
one can only obtain the direction of the Fermi velocity at nodes from the field-angle dependent quasiparticle excitation.
To reveal the gap structure in $k$-space, one should know the band structure of the material.
For these reasons, the gap-structure investigation of Q1D superconductors by the field-angle-induced quasiparticle excitation had not been explored.

\begin{figure}[htb]
\begin{center}
\includegraphics[width=0.9\hsize]{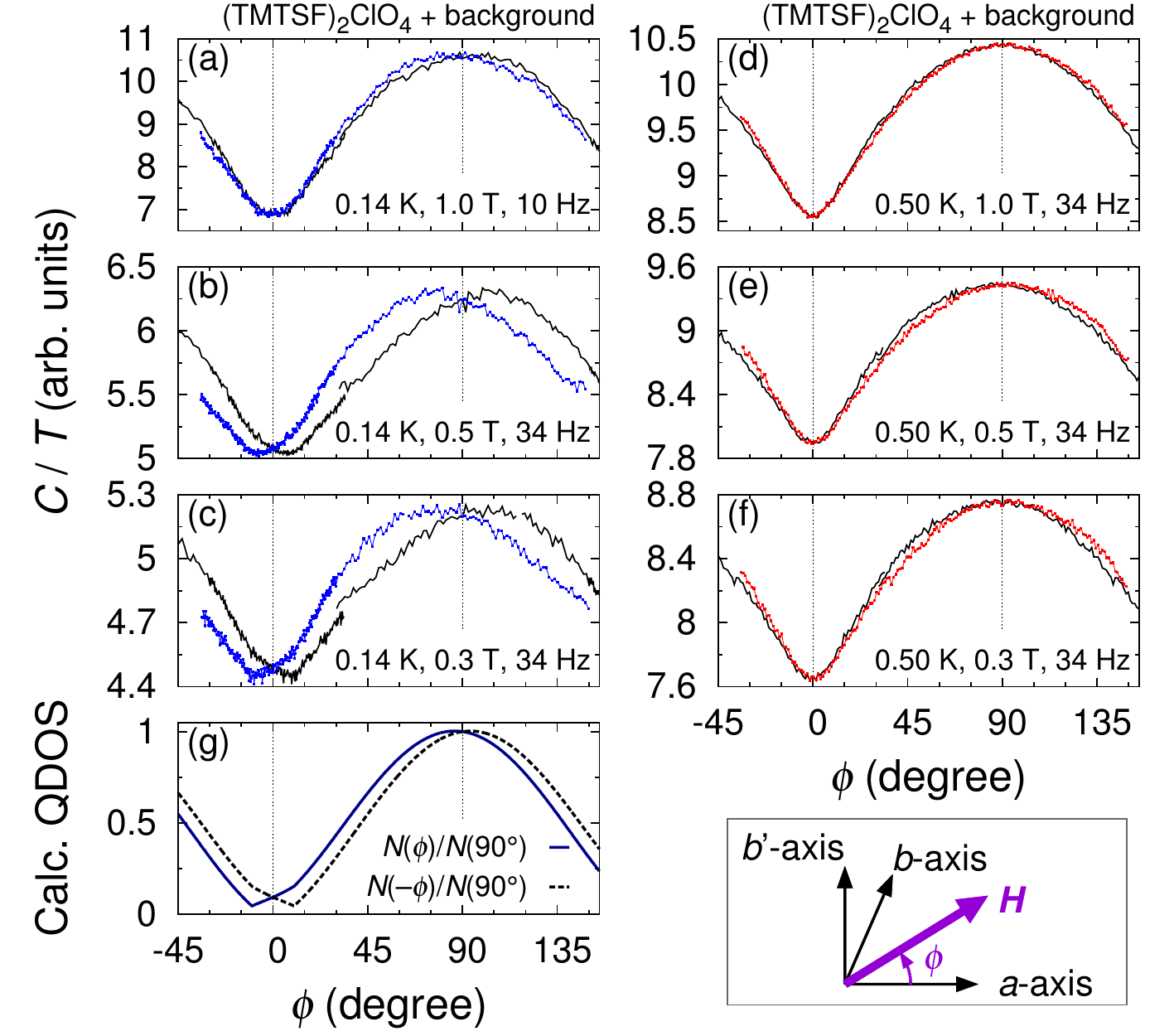}
\caption{(a)-(f) Observed in-plane field-angle dependence of the heat capacity of \tmcns~\cite{Yonezawa2012.PhysRevB.85.140502R}. Blue curves in panels (a)-(c) are $C/T$ obtained at 0.14~K and red curves in (d)-(f) are at 0.50~K. 
The black curves indicate $C/T$ plotted against $-\phi$. The difference between the colored and the black curves represents the asymmetry of the $C(\phi)/T$ curve. The curves in (g) are simulated results by calculating the density of states based on a simple Doppler-shift model with nodes at $\phi=\pm 10\deg$~\cite{Yonezawa2012.PhysRevB.85.140502R}. The definition of the in-plane field angle $\phi$ is indicated at the bottom-right corner.
\label{fig:Yonezawa2012_C-phi}
}
\end{center}
\end{figure}

Recently, such experiments was reported by Yonezawa \etal~\cite{Yonezawa2012.PhysRevB.85.140502R,Yonezawa2013.JPhysConfSer.449.012032},
who developed a highly sensitive calorimeter based on the ``bath modulating method''~\cite{Graebner1989} and measured the field-strength and field-angle dependence of the heat capacity of one single crystal of \tmcns.
The in-plane field-angle $\phi$ dependence presented in Fig.~\ref{fig:Yonezawa2012_C-phi} is of particular interest. 
In addition to the large oscillation in the heat capacity originating from the in-plane anisotropy of $\Hcc$, additional kink structures in $C(\phi)$ curves are observed.
The kinks are located at $\phi=\pm 10\deg$; i.e., for fields $\pm 10\deg$ away from the crystalline $a$ axis within the $ab$ plane.

\begin{figure}[htb]
\begin{center}
\includegraphics[width=0.7\hsize]{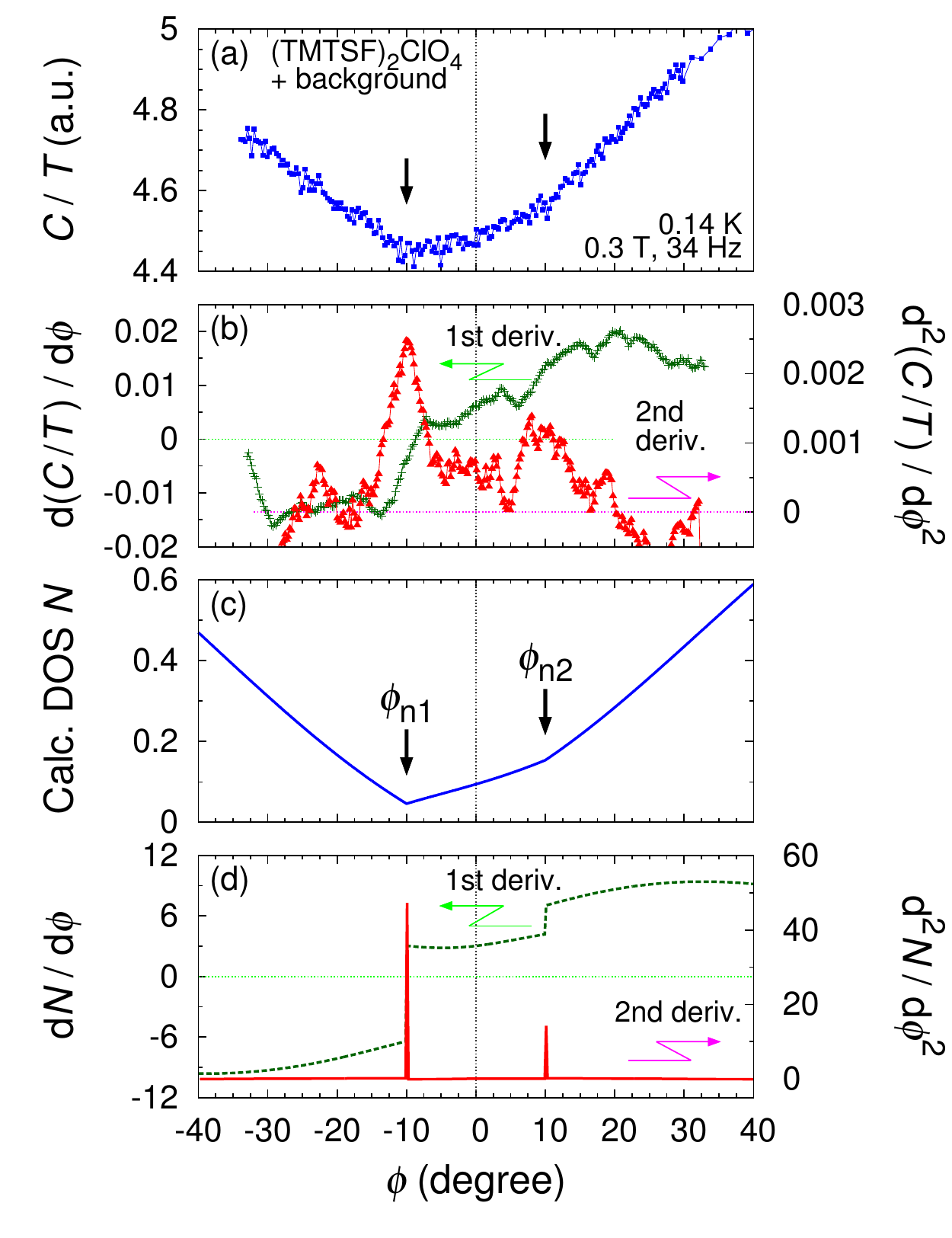}
\caption{ 
(a) In-plane field-angle dependence of the heat capacity of \tmc\ near $\phi=0\deg$~\cite{Yonezawa2012.PhysRevB.85.140502R}. The arrows indicate positions of the observed kinks. 
(b) First and second derivatives of $C(\phi)/T$. Anomalies at $\phi=\pm 10\deg$ corresponding to the kinks in $C(\phi)/T$ are easily seen. 
Calculated density of states $N$ and their derivatives based on a simple Doppler shift model with nodes at $\phi\sub{n1} = -10\deg$ and $\phi\sub{n2} = +10\deg$ are plotted in (c) and (d). \label{fig:Yonezawa2012_C-phi2}
}
\end{center}
\end{figure}

By comparing the experimental data with a simple simulation shown in Figs.~\ref{fig:Yonezawa2012_C-phi}(g), \ref{fig:Yonezawa2012_C-phi2}(c), and \ref{fig:Yonezawa2012_C-phi2}(d), it is suggested that $\pm 10\deg$ is the direction of $\vF$ at the gap nodes. 
Nagai \etal~\cite{Nagai2011.PhysRevB.83.104523} 
calculated the field-angle dependence of the specific heat based on the quasiclassical framework together with the first-principles band calculations, and they deduced similar conclusions.
Thus, it is now clarified that the Fermi-surface positions at which $\vF$ is pointing $\pm 10\deg$ away from the $a$ axis are candidate nodal positions in the $k$ space.

Based on the Fermi surface obtained by the tight-binding band calculation~\cite{Pevelen2001EurPhysJB}, Yonezawa \etal\ proposed that the $d$-wave-like state with nodes at $k_y=\pm 0.25/b^\ast$ 
best matches with 
experiment~\cite{Yonezawa2012.PhysRevB.85.140502R}. 
Here, $k_y$ is the 
wavevector perpendicular to the $a$ axis and $b^\ast =\pi/b $ is the size of the first Brillouin zone along the $k_y$ direction.
Considering the nesting vectors of the Fermi surface of \tmcns, this state is likely to be realized if the intra-band nesting plays the dominant role for Cooper pairing~\cite{Yonezawa2013.JPhysConfSer.449.012032}.

One comment should be made here; there still is a debate concerning the detailed Fermi-surface shape of \tmcns.
The conclusion of the nodal position in the $k$ space strongly depends on the assumption of the Fermi-surface shape, which is affected by the anion gap $\varDelta\sub{a}$.
A value of $\varDelta\sub{a}\sim 100$~meV has been used in the tight-binding-model calculation~\cite{Pevelen2001EurPhysJB}, and for the analysis of the specific-heat data~\cite{Yonezawa2012.PhysRevB.85.140502R}.
On the other hand, Nagai \etal~\cite{Nagai2011.PhysRevB.83.104523} recently performed first-principles band calculation for the anion-ordered low-temperature crystal structure of \tmcns,
and evaluated $\varDelta\sub{a}$ as nearly zero.
As a result, the calculated Fermi surfaces nearly touch each other.
Another first-principles calculation by Alemany \etal~\cite{Alemany2014.PhysRevB.89.155124} revealed small but sizable anion-order effect with $\varDelta\sub{a} \sim 14$~meV, accompanied by a weak anti-crossing between the split bands. 
Thus the resultant Fermi surfaces are well separated in the $k$ space.
More recent calculation by Aizawa \etal\ obtained a similar gap value $\varDelta\sub{a} \sim 8.7$~meV~\cite{Aizawa2015.unpublished}.
Experimentally, $\varDelta\sub{a}$ should be finite but seems to be no more than 25~meV~\cite{Uji1996.PhysRevB.53.14399,Yoshino1999.JPhysSocJpn.68.3142}.
A value around 14~meV is confirmed by a recent analysis of 
magnetoresistance oscillations in \tmc by G.~Montambaux and D.~Jerome~\cite{Montambaux2015}.

Returning back to the nodal SC gap structure, the $d$-wave-like state with nodes at $k_y=\pm 0.25b^\ast$ remains a candidate structure even with $\varDelta\sub{a}=0$~meV according to the detailed analysis~\cite{Yonezawa2013.JPhysConfSer.449.012032}, at least within the tight-binding model.
Experimental determination of $\varDelta\sub{a}$ and analyses based on the relevant band structure are
still necessary to resolve the nodal structure.
In addition, microscopic theories on the gap structure based on realistic band structures is also important to finally settle this issue of the exact nodal positions.

\section{High-field superconducting state}\label{sec:high-field-state}

As already mentioned in previous sections, \tms2x salts have been known to exhibit a divergent behavior of the transport $\Hcc$ with decreasing  temperatures.
The origin of this behavior has been attributed to spin-triplet pairing or to 
the Fulde-Ferrell-Larkin-Ovchinnikov (FFLO) states~\cite{Suzumura1998.ProgTheorPhys.70.654,Machida1984PhysRevB,Lebed1986.JETPLett.44.114,Dupuis1993,Dupuis1994,Miyazaki1999,Lebed1999,Lebed2000,Vaccarella2001.PhysRevB.63.180505R,Shimahara00,Belmechri2008.EurphysLett.82.47009,Aizawa2008.PhysRevB.77.144513,Aizawa2009.PhysRevLett.102.016403,Lebed2011.PhysRevLett.107.087004,Fuseya2012.EurophysLett.100.57008}.
In case of \tmc, the former is excluded, since a clear decrease in the spin susceptibility is observed~\cite{Shinagawa2007} as described in Sec.~\ref{sec:spin-susceptibility}.
In addition, sudden increase of the nuclear-lattice relaxation rate $1/T_1$ observed above around 2~T~\cite{Shinagawa2007} was considered as a consequence of the formation of unusual high-field SC phases.

\begin{figure}[htb]
\begin{center}
\includegraphics[width=0.75\hsize]{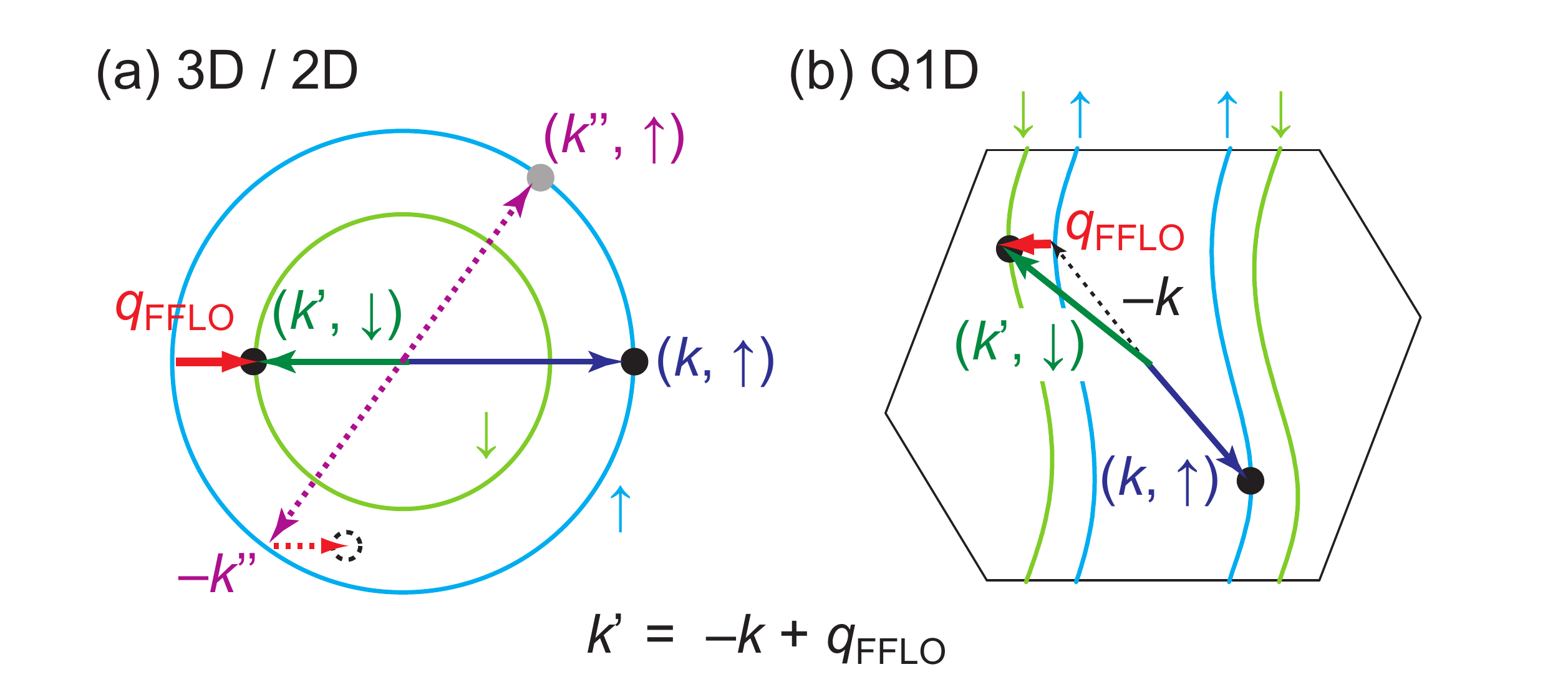}
\caption{Schematic comparison between FFLO pair formations for (a) 3D or 2D Fermi surfaces and (b) Q1D Fermi surfaces.
\label{fig:FFLO-3D-1D}
}
\end{center}
\end{figure}

The FFLO state~\cite{Fulde1964,Larkin1965} can be realized when spin-singlet Cooper pairs are formed among Zeeman-split Fermi surfaces in high magnetic fields~\cite{Matsuda2007JPhysSocJpnReview}.
Due to the Zeeman split, the Fermi wavenumber for the up-spin electron $\kFu$ and that for the down-spin electron $\kFd$ are not equal.
Thus, when a Cooper pair is formed between $\kFu$ and $-\kFd$ electrons as presented in Fig.~\ref{fig:FFLO-3D-1D}, the pair acquires the non-zero center-of-mass momentum $\qFFLO = \kFu-\kFd$. 
This momentum results in the spatial oscillation of the SC order parameter. 
This means that the FFLO state is accompanied by the translational symmetry breaking. 
In particular, for Q1D systems, $\qFFLO = \kFu-\kFd$ should be nearly fixed to the $a$ axis, since the number of pairs can be maximized if $\qFFLO$ matches with the nesting vector between the spin-up and spin-down Fermi surfaces, which is nearly parallel to the $a$ axis, as schematically shown in Fig.~\ref{fig:FFLO-3D-1D}(b).
Indeed, it is theoretically shown that the FFLO state with $\qFFLO \parallel a$ generally acquires high $\Tc$ in a Q1D system~\cite{Miyawaki2014.JPhysSocJpn.83.024703}.
Observation of unusual phenomena resulting from the symmetry breaking with such a fixed $\qFFLO$ can be a hallmark of the Q1D FFLO state.

There are only a few candidate materials for the FFLO state. The heavy Fermion compound CeCoIn\sub{5} clearly exhibits an unusual high-field phase~\cite{Radovan2003.Nature.425.51,Bianchi2003}.
However, this phase may not be a textbook-like FFLO state, since the phase is revealed to be accompanied by antiferromagnetic ordering~\cite{Kenzelmann2008.Science.321.1652}.
Other leading candidates are the two-dimensional organic superconductors $\kappa$-(BEDT-TTF)\sub{2}Cu(NCS)$_2$ and $\lambda$-(BETS)\sub{2}FeCl\sub{4}.
In the former, the existence of additional high-field SC phase has been confirmed by magnetic and thermodynamic measurements~\cite{Singleton2000.JPhysCondensMatter.12.L641,Lortz2007PhysRevLett}, as well as by an NMR study~\cite{Write2011.PhysRevLett.107.087002}.
More recently, substantial increase of $1/T_1$ attributable to the Andreev reflections originating from the order-parameter modulation is observed~\cite{Mayaffre2014.NaturePhys.10.928}.
In the latter compound, oscillatory behavior in the electric resistivity due to the vortex flow is observed~\cite{Uji2006.PhysRevLett.97.157001}. 
This behavior is believed to be a consequence of the ``locking'' effect between vortices and order-parameter modulation.
Its sister compound $\lambda$-(BETS)\sub{2}GaCl\sub{4} also exhibits a signature of the FFLO state~\cite{Tanatar2002PhysRevB}.

For \tmcns, only the unusual divergent-like behavior of $\Hcc(T)$ for $H\parallel b\dash$ had been known for the high-field state~\cite{Lee1995,Oh2004}.
In 2008, Yonezawa \etal\ investigated the in-plane field-angle dependence of the onset temperature of superconductivity, $\Tco$, based on the $\cstar$-axis resistance measurements of \tmcns\ single crystals~\cite{Yonezawa2008.PhysRevLett.100.117002,Yonezawa2008.JPhysSocJpn.77.054712}.
They made use of the anisotropy of $\Hcc$ to accurately deduce $\Tco$: they compared the resistance in fields exactly parallel to the $ab$ plane to those in fields tilted away from the $ab$ plane only by a few degrees.
The $\cstar$ axis component of the field induced by the tilting destroys the superconductivity, allowing one to extract contribution of superconductivity by comparing resistances for the two field directions.

\begin{figure}[htb]
\begin{center}
\includegraphics[width=1.0\hsize]{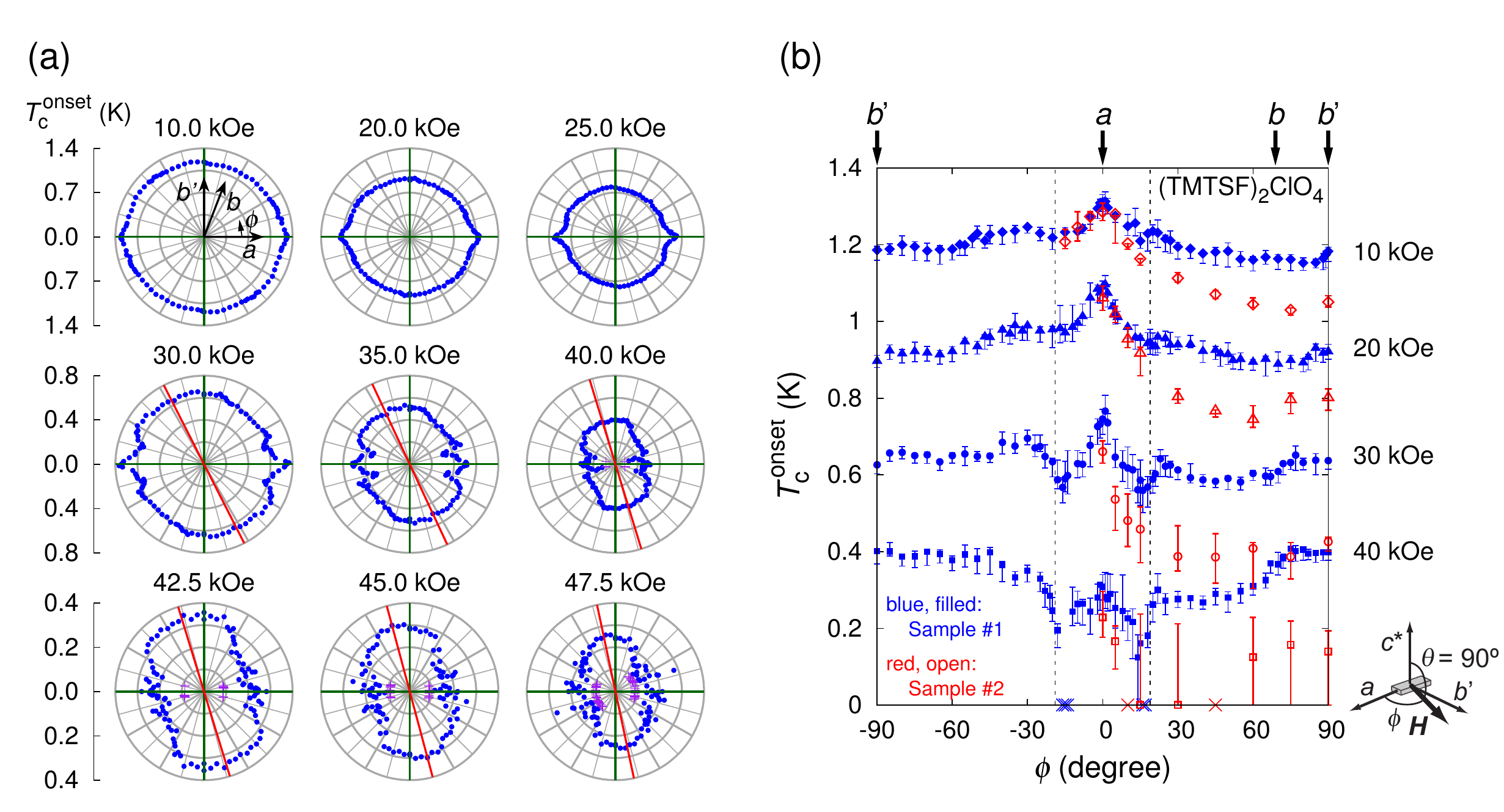}
\caption{(a) Polar plot of the $\phi$ dependence of $\Tco$ at several magnetic fields. The red line indicates the new principal axis emerging above 3~T~\cite{Yonezawa2008.PhysRevLett.100.117002}. 
(b) Comparison of $\Tco(\phi)$ for different samples~\cite{Yonezawa2008.JPhysSocJpn.77.054712}. The blue and red points indicate $\Tco$ of Sample~\#1 (very clean) and Sample~\#2 (moderately clean), respectively. Substantial difference is seen for $|\phi|>19\deg$, whereas the sample dependence is rather small for smaller field angles.
\label{fig:Yonezawa2008-Tc-phi}
This difference is attributed to the fact that the FFLO state, as well the field-induced 2D confinement for $H\parallel b\dash$, is very sensitive to impurity scatterings.
}
\end{center}
\end{figure}

It is revealed that, not only for $H\parallel b\dash$, but also for $H\parallel a$, $\Tco$ remains finite up to 5~T, the maximal field achieved in this study, as shown in Fig.~\ref{fig:phase-diagram}.
In particular, the onset curve for $H\parallel a$ exhibits a peculiar ``S'' shape, a limited behavior at around 0.8~K and an increase again below 0.3~K.
The behavior for $H\parallel a$ resembles that observed in the pressure-induced superconductivity in \tmpns~\cite{Lee1997}, and is recently theoretically treated within the FFLO scenario~\cite{Fuseya2012.EurophysLett.100.57008,Miyawaki2014.JPhysSocJpn.83.024703}.
What is more, unusual modulation in in-plane field-angle $\phi$ dependence of $\Tco$ is observed above 3.0~T.
In particular, the maxima of the $\Tco(\phi)$ curve, which is located at $\phi=0\deg$ ($H\parallel a$) and $\phi=90\deg$ ($H\parallel b\dash$) at low fields, the latter is found to shift away from the crystalline $b\dash$ axis at high fields.
This is in some sense a (quasi) field-induced breaking of symmetry.~\footnote{Since the crystal structure of \tmc\ belongs to the triclinic space group, any spatial symmetry except for the inversion symmetry is already broken by the lattice. Therefore, strictly speaking, it is not accurate to say ``symmetry breaking by the magnetic field'' in the present case. Nevertheless, $\Tco(\phi)$ exhibits nearly a mirror symmetry with respect to the $a$ and $b\dash$ axes in low field.}
To the best of our knowledge, such a modulation in $\Tc$ has never been reported in any other FFLO candidates. 
In addition, the high-field state is sensitively suppressed by a tiny amount of impurities~\cite{Yonezawa2008.JPhysSocJpn.77.054712} (Fig.~\ref{fig:Yonezawa2008-Tc-phi}(b).

This unusual phenomena is interpreted as a consequence of the formation of FFLO states.
In FFLO states, the modulation vector $\qFFLO$ of the SC order parameter breaks the translational symmetry of the SC state, and may lead to unusual field-angle dependence of $\Tco$. 
Such an interpretation has been indeed supported by recent theories.
Croitoru and Buzdin 
~\cite{Croitoru2012.PhysRevLett.108.207005,Croitoru2012.PhysRevB.86.224508} found that $\Tc(\phi)$ exhibit unusual $\phi$ dependence once the system is in the FFLO state, by solving linearized Eilenberger equations for $s$-wave superconductivity in a highly anisotropic quasi-two-dimensional (Q2D) model. 
More recently, they revealed similar results by quasiclassical formalisms for an $s$-wave model with a more-realistic Q1D band~\cite{Croitoru2014.PhysRevB.89.224506} and for a $d$-wave model but with a Q2D band~\cite{Croitoru2013.JPhysCondensMatter.25.125702}. 

\begin{figure}[htb]
\begin{center}
\includegraphics[width=1\hsize]{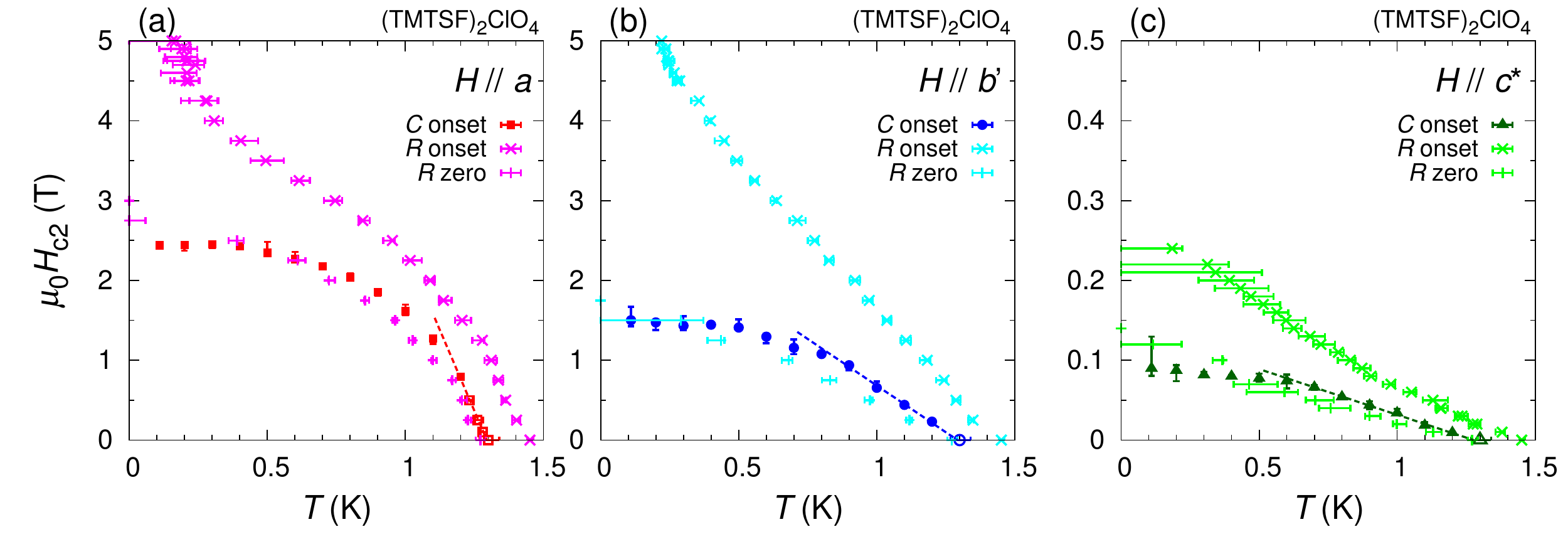}
\caption{SC phase diagram of \tmcns\ obtained by the specific heat (filled points) and resistivity measurements (crosses) for (a) $H\parallel a$, (b) $H\parallel b\dash$, and (c) $H\parallel \cstar$. Figures are made based on data in Refs.~\cite{Yonezawa2008.PhysRevLett.100.117002,Yonezawa2012.PhysRevB.85.140502R}. Notice that the vertical scale of the panel (c) is 20 time smaller than those of the other panels.
\label{fig:Yonezawa2012_phase-diagram}
}
\end{center}
\end{figure}

\begin{figure}[htb]
\begin{center}
\includegraphics[width=0.7\hsize]{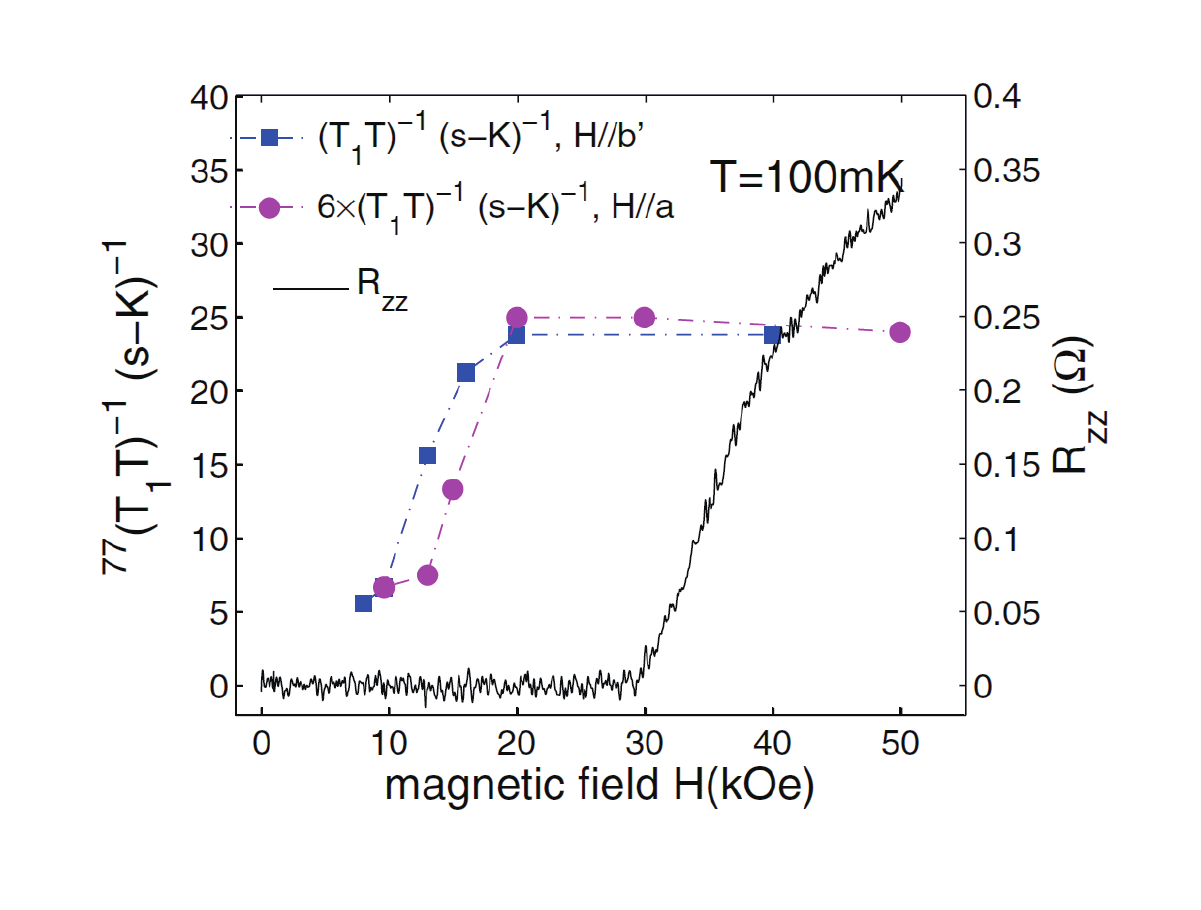}
\caption{Field dependence of $1/(T_1T)$ for $H\parallel a$ and $H\parallel b\dash$ obtained by the NMR study on \tmc~\cite{Shinagawa2007}. 
\label{fig:Shinagawa2007_T1-H}
}
\end{center}
\end{figure}

It is then natural that much effort has been devoted to search for thermodynamic evidence of the realization of the high-field FFLO phase.
Interestingly, a specific-heat study with accurately aligned magnetic fields revealed that an anomaly in the specific heat at $\Tco$ cannot be detected~\cite{Yonezawa2012.PhysRevB.85.140502R}:
For field directions along the three principal axes, the only detected anomaly is located close to the curve below which resistivity is zero as shown in Fig.~\ref{fig:Yonezawa2012_phase-diagram}.
(This may be just a coincident, since in some experiments the zero-resistance state is observed up to around 3~T for $H\parallel a$ and $H\parallel b\dash$~\cite{Oh2004, Shinagawa2007}. Also see Fig.~\ref{fig:Shinagawa2007_T1-H}.)
The field at which the specific heat anomaly is detected and the specific heat recovers its normal state value should be assigned as the thermodynamic $\Hcc$, and above this field  
superconductivity has a density of states nearly equal to that in the normal state.
Nevertheless, the resistivity anomaly observed above the thermodynamic $\Hcc$ is quite robust and has been reproduced by several groups~\cite{Lee1995,Oh2004,Yonezawa2008.PhysRevLett.100.117002}. 

One possible explanation is that the density of states in the high-field FFLO state is nearly equal to that in the normal state, because of the zero-gap region in real space originating from the order-parameter modulation~\cite{Fulde1964}.
This is reasonable, but experimental efforts to reveal the thermodynamic phase boundary between the FFLO state and normal state is highly required to support the scenario.
Another explanation is that the high-field FFLO region intrinsically acquires fluctuating nature, probably assisted by the low-dimensional electronic state in this material.
We emphasize here that, even with fluctuating superconductivity, the observed anomalous behavior in $\Tco(\phi)$ is rather difficult to be explained without (quasi) symmetry breaking in the underlying pairing channel.

\section{Metallic state above $\Tc$: antiferromagnetic fluctuation and its relation to superconductivity}

\begin{figure*}[t]	
\centerline{ \includegraphics[width=0.8\hsize]{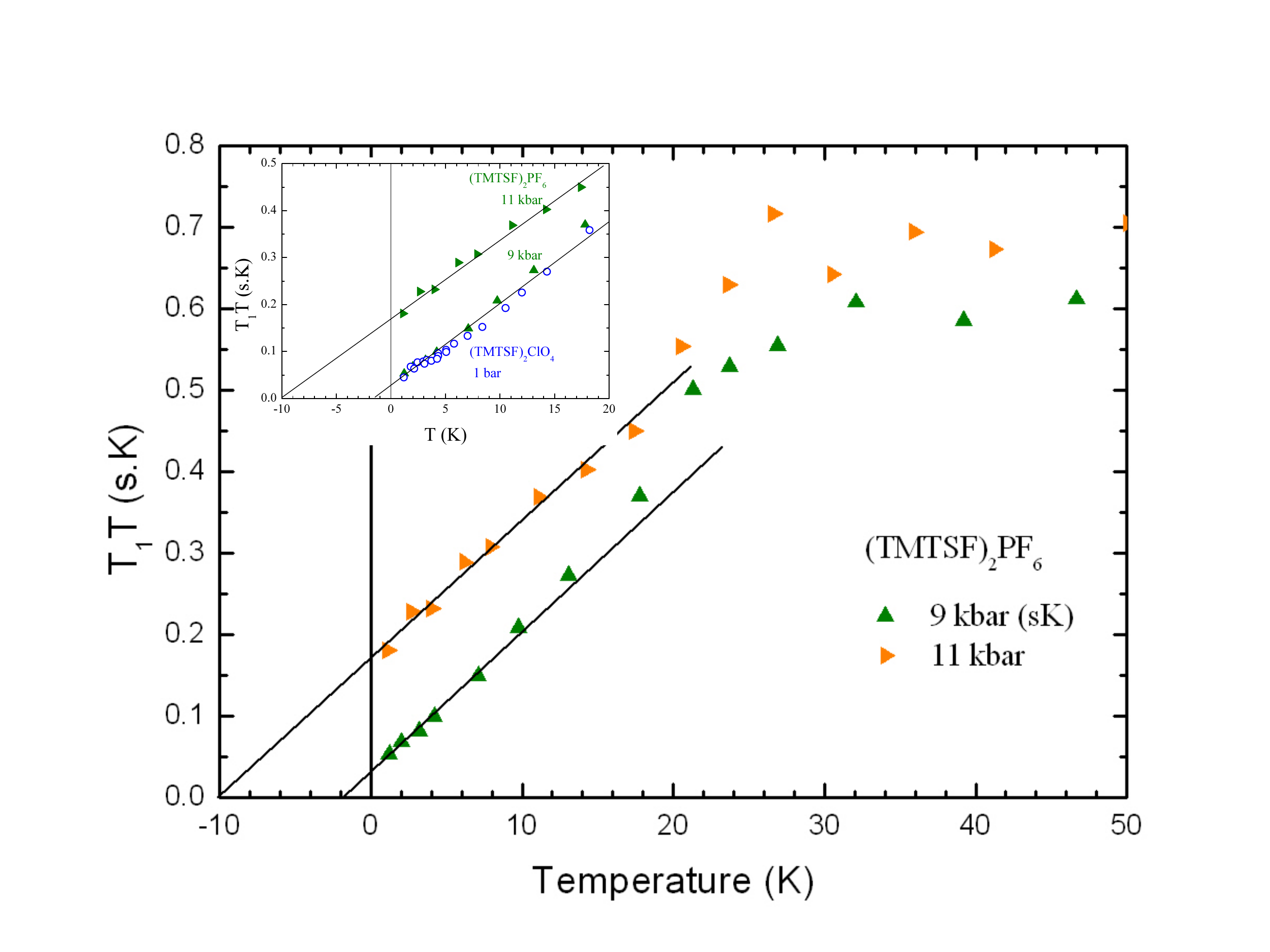}}
\caption{Temperature dependence of the nuclear relaxation time multiplied by temperature versus temperature according to the data of Ref.~\cite{Creuzet87}. A Korringa regime, $T_{1}T$ = const is observed down to $25$~K. The 2D AF regime is observed below $\approx 15$~K and the small Curie-Weiss temperature of the 9~kbar run is the signature of the contribution of  quantum critical fluctuations to the nuclear relaxation. The Curie-Weiss temperature becomes zero at the QCP. These data show that the QCP should be slightly below 9~kbar with the present pressure scale. The inset shows that the organic superconductor \tmc at ambient pressure is very close to fulfill  quantum critical conditions.}
\label{T1TvsT_PF6ClO4} 
\end{figure*}
\begin{figure*}[htbp]	
\centerline{ \includegraphics[width=0.7\hsize]{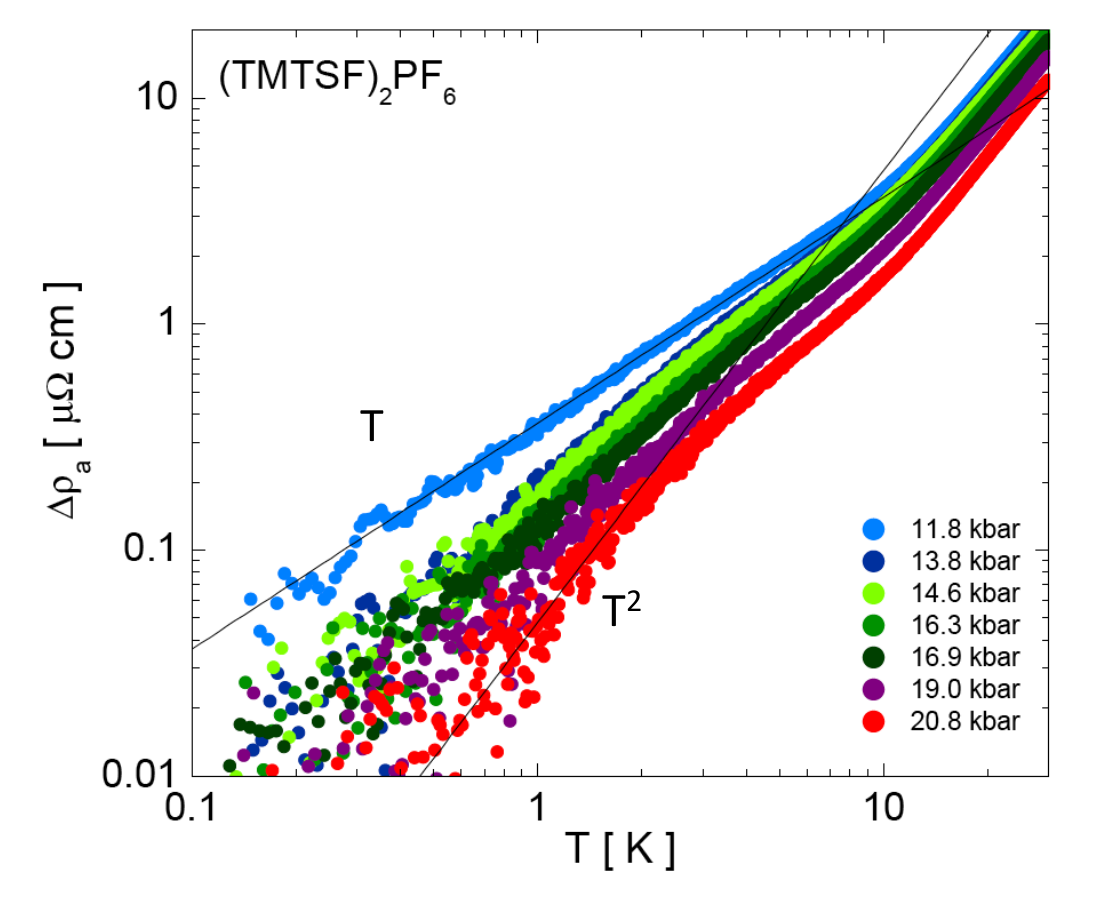}}
\caption{A log-log plot of the inelastic  longitudinal resistivity of \tmp6  below $20$~K, according to Ref.~\cite{Doiron09}.}
\label{figureTandT2}  
\end{figure*}

Interestingly, the metallic phase of \tmpns\ in the 3D coherent regime when pressure is in the neighbourhood of the critical pressure \pc behaves in a way far from what is expected for a Fermi liquid. 
This behavior indicates the dominance of quantum critical fluctuations near \pcns.
Moreover, close relation between the non-Fermi-liquid behavior and superconductivity has been recently revealed both experimentally and theoretically, as described in detail below.

Experimentally, NMR measurements of $1/T_1$ have probed antiferromagnetic fluctuations.
The canonical Korringa law, $1/T_{1}T \propto {\chi^{2} (q=0,T)} $, is well obeyed at high temperatures, say, above 25 K, but the low temperature behaviour deviates strongly from the standard relaxation in paramagnetic metals. As shown in Fig.~\ref{T1TvsT_PF6ClO4}, an additional contribution to the relaxation rate emerges on top of the usual Korringa relaxation. This additional contribution rising at low temperatures has been attributed to the onset of antiferromagnetic fluctuations in the vicinity of \pc~\cite{Doiron09,Doiron10,Bourbonnais11}. On the other hand, in the lower temperature regime, the relaxation rate follows a law such that $T_{1}T=C(T+\varTheta)$ as shown in Fig.~\ref{T1TvsT_PF6ClO4}. This is the Curie-Weiss behaviour for the relaxation which is to be observed in a 2D fluctuating antiferromagnet~\cite{Brown08,Wu05,Moriya00,Bourbonnais09}.
Similar behavior is also found in a ${}^{13}$C NMR study~\cite{Kimura2011.PhysRevB.84.045123}.

The positive Curie-Weiss temperature $\varTheta$, which provides the energy scale of the fluctuations, becomes zero when pressure is equal to \pc (the quantum critical conditions). When $\varTheta$ becomes large comparable to $T$, the standard relaxation mechanism is expected to recover down to low temperatures, in agreement with the observation at very high pressures~\cite{Wzietek93}. 

The existence of fluctuations is also observed as anomalous behavior in transport.
At $P=P\sub{c}$, the inelastic scattering in transport reveals at once a strong linear term at low temperatures, as presented in a log-log plot of the resistivity versus $T$, Fig.~\ref{figureTandT2}.
This strongly linear behavior evolves to quadratic behavior in the high temperature regime. 
As pressure is increased away from $P\sub{c}$, the resistivity exhibits a general tendency to become quadratic at all temperatures~\cite{Doiron09} (see Figs.~\ref{figureTandT2} and \ref{molecule,structure,supra2}).
The existence of a linear temperature dependence of the resistivity is \textit{at variance} with the $T^2$ dependence expected from the ordinary electron-electron scattering in a conventional Fermi liquid, indicating that the dominant scattering involves spin fluctuations.

\begin{figure}[htb]	
\centerline{ \includegraphics[width=0.7\hsize]{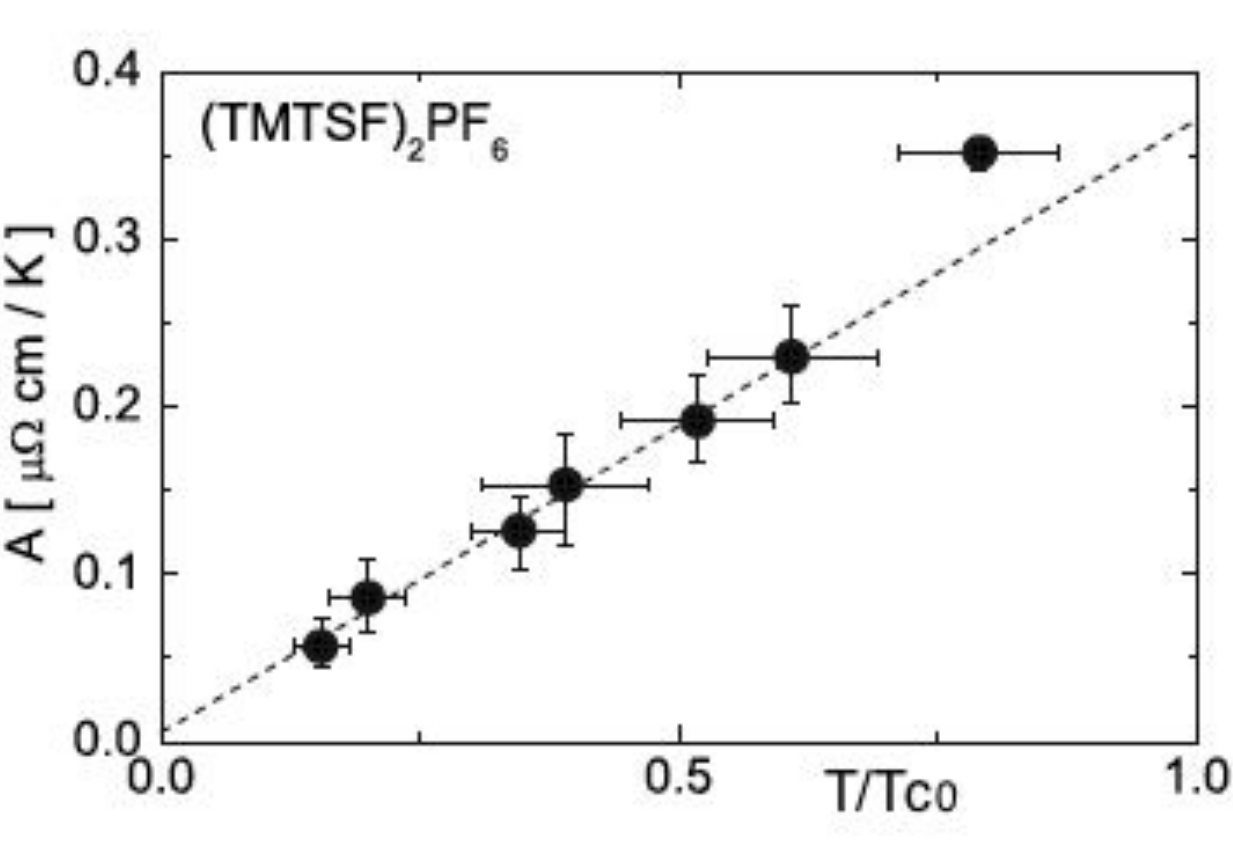}}
\caption{Coefficient $A$ of linear resistivity as a function of $\Tc$ plotted versus $\Tc/\Tcz$  for \tmpns. $\Tc$ is defined as the midpoint of the transition and the error bars come from the 10\% and 90\% points, and $\Tcz$ is defined as $\Tcz=1.23$~K, the maximal $\Tc$ under the pressure of 8~kbar in the SDW/SC coexistence regime. The dashed line is a linear fit to all data points excluding that at $\Tc = 0.87$~K, according to Ref.~\cite{Doiron09}.}
\label{AvsTcPF6.pdf} 
\end{figure}

Furthermore, the investigation of both transport and superconductivity under pressure in \tmpns\ has established a correlation between  the amplitude 
of   the linear temperature dependence of the resistivity  and the value of $\Tc$, as displayed in Fig.~\ref{AvsTcPF6.pdf}. 
This correlation suggests a common origin for  the inelastic scattering of the metallic phase and pairing in the SC phase \tmpns~\cite{Doiron09}, as discussed in the rest of this section.

Within the framework of a weak-coupling limit, the problem of the interplay between antiferromagnetism and superconductivity in the Bechgaard salts has been theoretically worked out using the renormalization group (RG) approach~\cite{Bourbonnais09,Nickel05} as summarized below.
The theories take into account only the 2D problem.  
The RG integration of high-energy electronic degrees of freedom was  carried out  down to the Fermi level, and leads to a  renormalization of the couplings at  the temperature $T$~\cite{Duprat01,Nickel06,Bourbonnais09}.  The RG flow   superimposes  the $2k_F$ electron-hole (density-wave) and Cooper pairing many-body processes, which combine and interfere at every order of perturbation. As a function of the `pressure' parameter $t_{b}'$, \textit{i.e} the unnesting interchain coupling, a  singularity in the scattering amplitudes signals an instability of the metallic state toward the formation of an ordered state at some characteristic temperature scale. At low $t_{b}'$, nesting is sufficiently strong to induce a SDW instability in  the temperature range  of experimentally observed $T_{\rm SDW}\sim 10$-20~K. 


When the antinesting parameter approaches the threshold coupling $t_{b}'^*$  from below  ($t_{b}'^* \approx 25.4~{\rm K}$ using the above  parameters), $T_{\rm SDW}$ sharply decreases as a result of interference  between the Cooper  and the Peierls channel  (SDW correlations). This situation  leads in turn to an attractive pairing  in the SC $d$-wave (SCd) channel. 
This gives rise to an instability of the normal state against SCd order at the temperature $\Tc$ with pairing coming from antiferromagnetic spin fluctuations between carriers of neighbouring chains. 
Such a pairing model actually supports the conjecture of interchain pairing in order for the electrons to avoid the Coulomb repulsion made  by V.~Emery in 1983 and 1986~\cite{Emery83,Emery86}.

\begin{figure*}[htbp]	
\centerline{ \includegraphics[width=0.6\hsize]{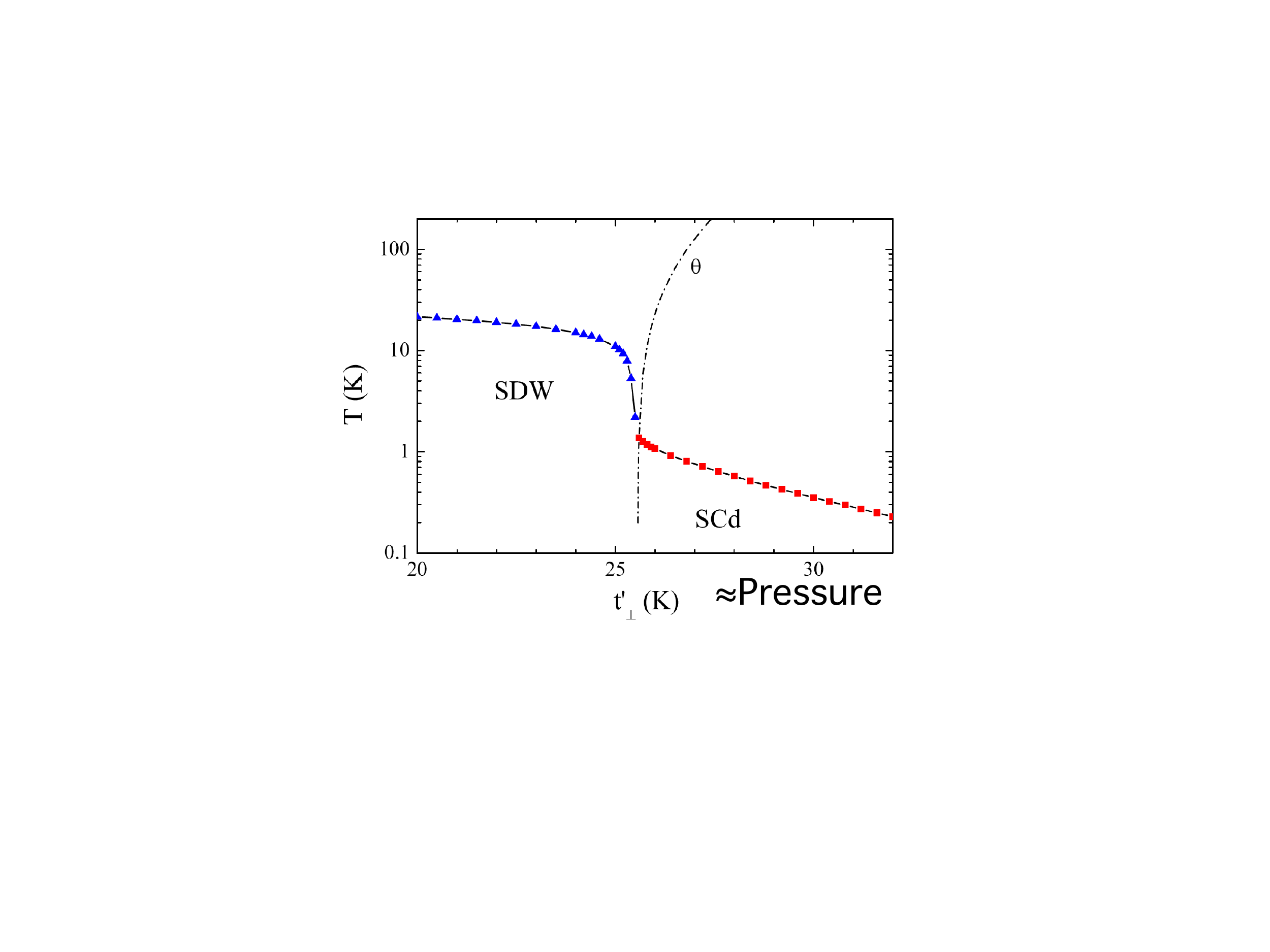}}
\caption{Calculated phase diagram of the quasi-one-dimensional electron gas model from the renormalization group method at the one-loop level~\cite{Bourbonnais09}.
$\varTheta$ and 
the dash-dotted line defines the temperature region of the Curie-Weiss behavior for the inverse
normalized SDW response function.
}
\label{Diagtheo} 
\end{figure*}

The calculated phase diagram shown in Fig.~\ref{Diagtheo} with reasonable parameters $g_1=g_2/2 \approx 0.32$ for the backward and forward scattering amplitudes respectively and $g_3\approx 0.02$ for the longitudinal  Umklapp scattering term~\cite{Bourbonnais09,Bourbonnais11} captures the essential features of the experimentally-determined phase diagram of \tmpns\ presented in Fig.~\ref{molecule,structure,supra2}.

Sedeki \etal~\cite{Sedeki10} have proceeded to an  evaluation of the imaginary part of the one-particle self-energy.  In addition to the regular Fermi-liquid component, whose scattering rate goes as $T^2$, low-frequency spin fluctuations yield $\tau^{-1} = aT\xi $, where $a$ is a constant and the antiferromagnetic correlation length $\xi(T)$ increases according to $\xi = c(T + \varTheta)^{-1/2}$ as $T \rightarrow \Tc$, where $\varTheta$ is the temperature scale for spin fluctuations~\cite{Sedeki10}. It is then natural to expect the Umklapp resistivity to contain  (in the limit $T \ll \varTheta$) a linear term $AT$, whose magnitude would presumably be correlated with $\Tc$, as both scattering and pairing are caused by the same antiferromagnetic correlations. The observation of a $T$-linear law for the resistivity up to 8~K  in \tmpns\ under a pressure of 11.8 kbar as displayed in Fig.~\ref{figureTandT2} is therefore consistent with the value of $\varTheta=8$~K determined from NMR relaxation at 11~kbar displayed in Fig.~\ref{T1TvsT_PF6ClO4}.
More recently, Bakrim and Bourbonnais~\cite{Bakrim2014.PhysRevB.90.125119} studied the effects of electron-phonon interactions on the SC and SDW channels. Interestingly, it is revealed that electron-phonon coupling enhances spin fluctuation, leading to unusual phenomena such as the positive isotope effect.

We add one comment that, in \tms2x, the existence of the quantum critical point is actually not trivial because the boundary between the SDW and SC phases is a first-order phase transition within the pressure-temperature phase diagram, in contrast to ordinary theories on quantum criticality assuming a second-order transition.
However, it has been recently revealed that other typical ``quantum critical'' materials such as iron pnictides~\cite{Goko2009.PhysRevB.80.024508} indeed exhibit first-order-like behavior in the vicinity of the quantum critical point, evidenced by phase separation between magnetically ordered and paramagnetic phases detected by $\muup$SR studies~\cite{Uemura2015.Book}.
Thus, it is now getting clearer that the quantum criticality near a first-order transition observed in \tms2x probably shares general and important physics with a broad class of materials.

\section{Conclusion}

Both experimental and theoretical results point to the contribution of electron correlations to the SC pairing problem. The extensive experimental evidence  in favor of  the   emergence of superconductivity in the \tm2x family next to the  stability pressure threshold for antiferromagnetism  has shown the need for a unified description of all electronic excitations that lies at the core of  both density-wave and SC correlations. In this matter, the  recent progresses of the renormalization group method for the 1D-2D electron  gas model  have resulted  in  predictions about the possible symmetries of the SC order parameter when a purely electronic mechanism is involved,  predictions that often differ from  phenomenologically based approaches to superconductivity but are in fair agreement with  recent experimental findings.  

To summarize, firstly, the SC order parameter is displaying lines of nodes which are governing the stability against impurity and thermodynamics of the SC phase. Important constraints on the nodal position have been obtained by the field angular dependence of the specific heat.
Secondly, the electron scattering in the metallic phase above \tc suggests the existence of  strong antiferromagnetic fluctuations leading to the possibility of a spin mediated pairing in the SC phase. The pairing mechanism behind organic superconductivity is likely different from the proposal made by Little but it is nevertheless a phonon-less mechanism, at least in \tm2x\ superconductors.

What is also emerging from the work on these prototype 1D  organic superconductors is their very simple electronic nature  with only a single band at Fermi level, no prominent spin orbit coupling and extremely high chemical purity and stability. 
They should be considered in several respects as model systems to inspire the physics of the more complex high \tc superconductors, especially for pnictides and electron-doped cuprates. 
Most concepts discovered in these simple low dimensional conductors may also become of interest for the study of other 1D or Q1D systems such as carbon nanotubes, artificial 1D structures, the purple bronze superconductor Li\sub{0.9}Mo\sub{6}O\sub{17} with Mo-O chains~\cite{Greenblatt1984,Xu2009.PhysRevLett.102.206602}, the newly-discovered telluride superconductor Ta\sub{4}Pd\sub{3}Te\sub{16} with Ta-Pd chains~\cite{Jiao2014.JAmChemSoc.136.1284}, and the recently discovered $A_{2}$Cr\sub{3}As\sub{3} ($A={}$ K, Rb, Cs) materials comprising [(Cr\sub{3}As\sub{3})$^{2-}$]$_{\infty}$ chains~\cite{Tang2015.PhysRevB.91.020506R,Bao2015.PhysRevX.5.011013}.
It should be noted that the electronic anisotropy of the latter two classes of compounds seems to be weaker than originally expected~\cite{Singh2014.PhysRevB.90.144501,Kong2015.PhysRevB.91.020507R} and much weaker than those of the Bechgaard superconductors.
Nevertheless, unconventional behavior, such as possible nodal superconductivity in Ta\sub{4}Pd\sub{3}Te\sub{16}~\cite{Pan2014.arXiv.1404.0371} and unusually large $\Hcc$ in Li\sub{0.9}Mo\sub{6}O\sub{17}~\cite{Mercure2012.PhysRevLett.108.187003} and $A_{2}$Cr\sub{3}As\sub{3}~\cite{Tang2015.PhysRevB.91.020506R,Bao2015.PhysRevX.5.011013,Kong2015.PhysRevB.91.020507R}, resembles those observed in \tms2x and thus it is interesting to explore the common nature of Q1D superconductivity amongst a wide class of materials.
Of course, serious work using clean single crystals should be accomplished before truly establishing any 1D/Q1D physics governing SC properties.
This is actually what has been done on the Bechgaard salts for more than 30 years.

This article shows that there is still plenty of food for 
thought in the field of organic superconductors.

\section*{Acknowledgements}

We are grateful to Professor Jacques Friedel who has welcomed and strongly supported the research activity on low dimensional conductors at Orsay from its very beginning. He has contributed through  continuous encouragements and numerous  discussions. 
D.J. wishes to acknowledge the remarkably fruitful cooperation with Klaus Bechgaard who provided the samples for the experiments performed in Kyoto, Patrick Batail for the chemistry of various 1D and 2D conductors, with late Heinz Schulz, Thierry Giamarchi and Claude Bourbonnais for the theory,  with the group of Louis~Taillefer at Sherbrooke for recent experimental work, with Stuart Brown at UCLA and with our Orsay colleagues C. Pasquier, N. Joo  and P. Senzier.
S.Y. acknowledges Y.~Maeno for his great support and guidance, K.~Ishida, H.~Aizawa, K.~Kuroki for useful discussions, and T.~Kajikawa, S.~Kusaba, for technical assistance,
and M.~Oudah and I.~Kostylev for improving the text.
This work, as well as some studies explained in the text, has been supported in France by CNRS; and in Japan by Grants-in-Aids for Scientific Research on Innovative Areas on ``Molecular Degree of Freedom'' (KAKENHI 21110516, 23110715), ``Topological Quantum Phenomena'' (KAKENHI 22103002) and ``Topological Materials Science''  (KAKENHI 15H05852) from Ministry of Education, Culture, Sports, Science and Technology (MEXT) of Japan; and Grants-in-Aids for Scientific Research (KAKENHI 21740253, 23540407, 23110715, and 26287078)  from Japan Society for the Promotion of Science (JSPS).

\pagebreak


\bibliographystyle{./elsarticle-num_notitle_nonumber.bst}

\bibliography{CR_v7-4}



\end{document}